\documentclass[prd,aps,twocolumn,nofootinbib]{revtex4-1}
\usepackage{mathbbold}
\usepackage{amsfonts}
\usepackage{amsmath}
\usepackage{graphicx}
\usepackage{subfig}

\usepackage{fancyhdr}
\usepackage[marginal]{footmisc}

\usepackage{booktabs,dcolumn}

\usepackage{color}

\def \D {{\rm d}}

\begin{document}
\title{First stage of LISA data processing: Clock
synchronization and arm-length determination via a hybrid-extended Kalman filter}
\author{Yan Wang}
\email{yan.wang@aei.mpg.de}
\affiliation{Max-Planck-Institut f\"ur Gravitationsphysik (Albert-Einstein-Institut), Callinstra{\ss}e 38, 30167 Hannover, Germany}
\author{Gerhard Heinzel}
\email{ghh@mpq.mpg.de}
\affiliation{Max-Planck-Institut f\"ur Gravitationsphysik (Albert-Einstein-Institut), Callinstra{\ss}e 38, 30167 Hannover, Germany}
\author{Karsten Danzmann}
\email{karsten.danzmann@aei.mpg.de}
\affiliation{Max-Planck-Institut f\"ur Gravitationsphysik (Albert-Einstein-Institut), Callinstra{\ss}e 38, 30167 Hannover, Germany}
\begin{abstract}
In this paper, we describe a hybrid-extended Kalman filter algorithm to synchronize the clocks and to precisely determine the
inter-spacecraft distances for space-based gravitational wave detectors, such as (e)LISA. According to the simulation, the algorithm
has significantly improved the ranging accuracy and synchronized the clocks, making the phase-meter raw measurements qualified
for time-delay interferometry algorithms.
\end{abstract}
\maketitle

\section{Introduction}

The Laser Interferometer Space Antenna (LISA) \cite{LISA98,Danzmann03,LISA11} is a space-borne gravitational wave (GW) detector, aimed at various kinds of GW
signals in the low-frequency band between $0.1\,$mHz and $1\,$Hz. It consists of three identical spacecraft (S/C), each individually following a slightly elliptical
orbit around the Sun, trailing the Earth by about $20^\circ$. These orbits are chosen such that the three S/C retain an equilateral triangular configuration with an arm length of about $5\times10^9\,$m as much as possible. This is accomplished by tilting the plane of the triangle by about $60^\circ$ out of the ecliptic.
Graphically, the triangular configuration makes a cartwheel motion around the Sun. As a (evolving) variation of LISA, eLISA \cite{Danzmann13} is an ESA L2/L3 candidate space-based GW detector. It consists of one mother S/C and two daughter S/C, separated from each other by $1\times10^9\,$m. Although the configurations are slightly different, the principles and the techniques are equally applicable. Therefore, we will mainly focus on LISA hereafter.

\begin{figure}
\includegraphics[width=0.5\textwidth]{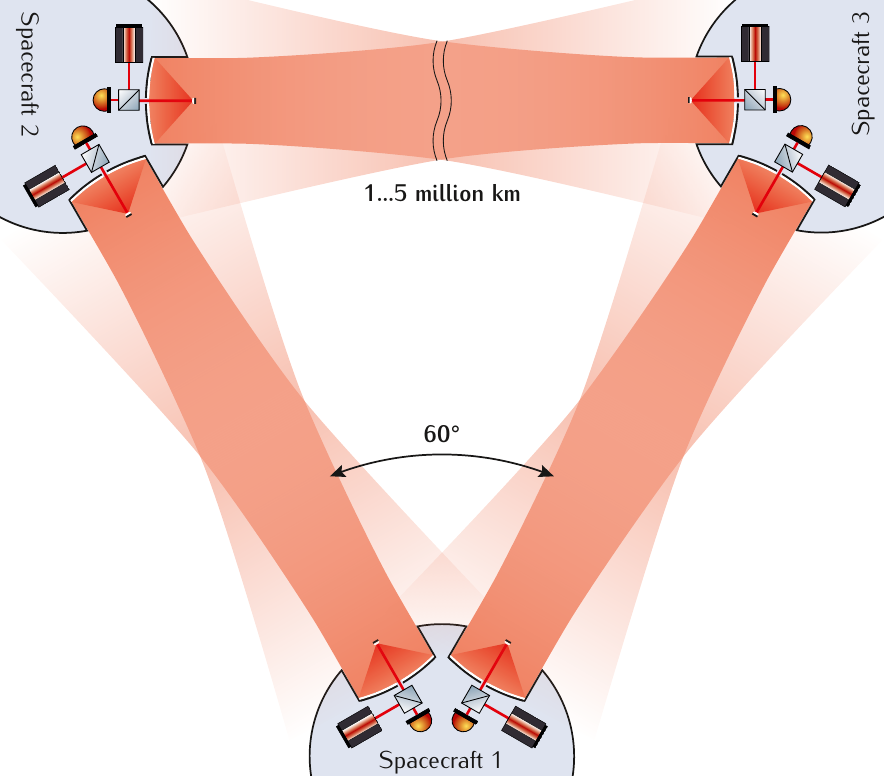}
\caption{ \label{fig:LISA_const} Schematic configuration of LISA S/C and the exchanged laser beams (by S. Barke \cite{LISAreport}).}
\end{figure}

Since GWs are propagating spacetime perturbations, they induce proper distance variations between test masses (TMs) \cite{Carbone}, which are free-falling references inside the S/C shield. LISA measures GW signals by monitoring distance changes between S/C. Spacetime is very stiff. Usually, even a fairly strong GW still produces spacetime perturbations only of order about $10^{-21}$ in dimensionless strain. This strain amplitude can introduce distance changes at the pm level in a $5\times10^9\,$m arm length. Therefore, a capable GW detector must be able to monitor distance changes with this accuracy. The extremely precise measurements are supposed to be achieved by large laser interferometers. A schematic classic LISA configuration with exchanged laser beams is shown in Fig.~\ref{fig:LISA_const}. LISA makes use of heterodyne interferometers with coherent offset-phase locked transponders \cite{McNamara05}. The phasemeter \cite{Shaddock06} measurements at each end are combined in postprocessing to form the equivalent of one or more Michelson interferometers. Information of proper-distance variations between TMs is contained in the phasemeter measurements.

Unlike the several existing ground-based interferometric GW detectors~\cite{LIGO92,GEO02,VIRGO97}, the armlengths of LISA are varying significantly with time due to celestial mechanics in the solar system. As a result, the arm lengths differ by about $1\%$ ($5\times10^7\,$m), and the dominating laser frequency noise will not cancel out. The remaining laser frequency noise would be stronger than other noises by many orders of magnitude. Fortunately, the coupling between distance variations and the laser frequency noise is very well known and understood. Therefore, we can use time-delay interferometry (TDI) techniques~\cite{TDIexperiment,TDI03,TDI99,TDI03b,TDI04,TDI05,TDI02,TDI05b,TDI12}, which combine the measurement data series with appropriate time delays, in order to cancel the laser frequency noise to the desired level.

However, the performance of TDI \cite{Petiteau08,TDI05} depends largely on the knowledge of armlengths and relative longitudinal velocities between the S/C, which are required to determine the correct delays to be adopted in the TDI combinations. In addition, the raw data are referred to the individual spacecraft clocks, which are not physically synchronized but independently drifting and jittering. This timing mismatch would degrade the performance of TDI variables. Therefore, they need to be referred to a virtual common ``constellation clock" which needs to be synthesized from the inter-spacecraft measurements. Simultaneously, one also needs to extract the inter-spacecraft separations and synchronize the time-stamps properly to ensure the TDI performance. More precisely, the knowledge of the distances between S/C needs to be better than 1 m rms at 3 Hz. Accordingly, the differential clock errors between the S/C are required to be estimated to a precision of 3.3 ns rms at 3 Hz\footnote{ Better knowledge of the armlengths and the differential clock errors will result in better cancellation of laser frequency noise in TDI variables, since the residual laser frequency noise in TDI variables is proportional to the armlength errors and the differential clock errors.}.  These are the main goal of the first stage of LISA data processing, which is the main topic of this paper.

The paper is organized as follows. In the next section, we will describe the entire LISA data processing chain, identifying the first stage of LISA data analysis.
In Sec.~\ref{sec:Meas1} and Sec.~\ref{sec:Meas2}, we will introduce and formulate the inter-spacecraft measurements. In Sec.~\ref{sec:Kalman} and \ref{sec:LISAmodel},
we describe the hybrid-extended Kalman filter algorithm and design a Kalman filter model for LISA. In Sec.~\ref{sec:Simulation}, we show the simulation results.
Finally the summary comes in Sec.~\ref{sec:Sum}.

\section{Overview of the entire LISA data processing chain}
\label{sec:LISAchain}

\begin{figure}
\includegraphics[width=0.5\textwidth]{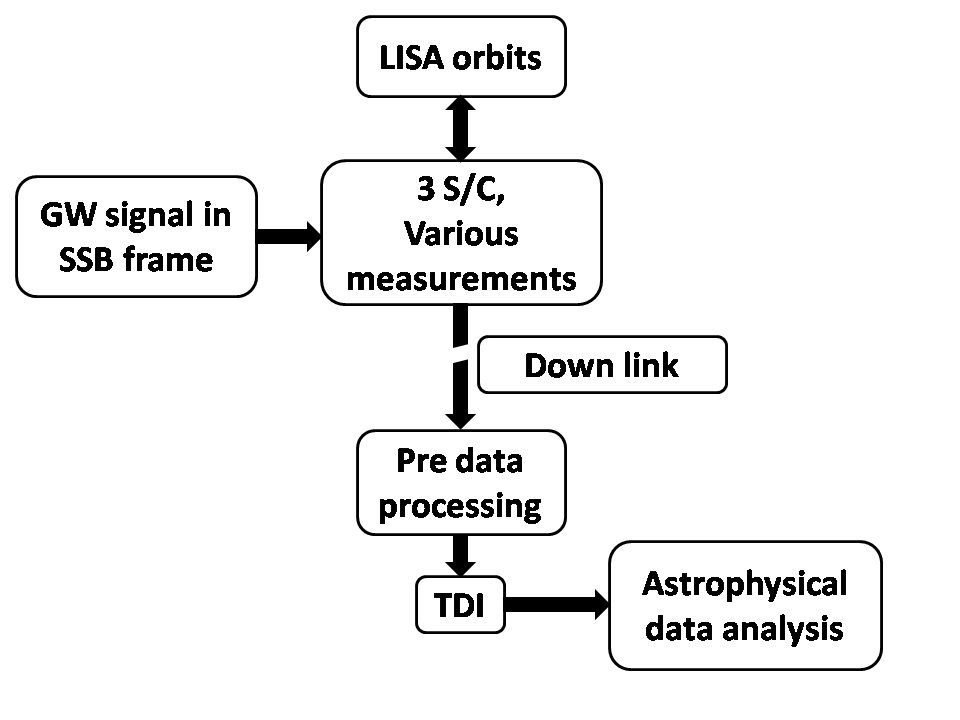}
\caption{ \label{fig:LISA_DAchain} LISA data processing chain.}
\end{figure}

In this section, we will talk about the perspective of a complete LISA simulation. The future goal is to simulate the entire LISA data processing chain
as detailed as one can, so that one will be able to test the fidelity of the LISA data processing chain, verify the science potential of LISA and set requirements for
the instruments. The flow chart of the whole simulation is shown in Fig.~\ref{fig:LISA_DAchain}.

The first step is to simulate LISA orbits \cite{Danzmann03,Li08} under the solar system dynamics. It should provide the position and velocity of each TM, or roughly S/C, as functions of some nominal time, e.g. UTC (Coordinated universal time), for subsequent simulations. Since TDI requires knowledge of the delayed armlengths (or light travel time) down to meter accuracies \cite{LISA11}, and the pre data processing algorithms could hopefully determine the delayed armlengths to centimeter accuracies, the position information provided should be more accurate than centimeters. In this paper, we will adopt Kepler orbits in the simulation.

The second step is to simulate GWs. There are various kinds of GW sources \cite{LISA11,Babak08} in the LISA band, such as massive black hole (MBH) binaries, extreme-mass-ratio inspirals (EMRIs), intermediate-mass-ratio inspirals (IMRIs), galactic white dwarf binaries (WDBs), gravitational wave cosmic background etc.
However, the simulation of GWs is irrelevant to the pre-processing simulation in this paper, since GWs introduce armlength variations to LISA at the pm level, which is many orders of magnitude below the ranging accuracy considered in the pre-processing stage.

The third step is to simulate the measurements and the noise.
%
The most relevant measurements to this paper are science measurements, ranging measurements, clock side band beat-notes\footnote{There are many more measurements, such as S/C positions and clock offsets observed by deep space network (DSN), various auxiliary measurements, incident beam angle measured by differential wavefront sensing (DWS).}.
Meanwhile, there are various kinds of noise sources \cite{Bender03,Stebbins04,Folkner97}, such as the laser-frequency noise, clock errors, the readout noise, the acceleration noise. Since the ranging and timing problem to be solved in the pre-processing stage in this paper is at millimeter to meter level, only the laser-frequency noise and the clock errors are relevant. See more discussions of these measurements and noise in Sec.~\ref{sec:Meas1}, \ref{sec:Meas2} and \ref{sec:Simulation}.

The ``down link" is referred to as a procedure of transferring the onboard measurement data back to Earth, which is also an important step in the simulation.
Since the beat notes between the incoming laser beam and the local laser are in the MHz range, the sampling rate of analog-to-digital convertors (ADCs) should be at least twice that, i.e.\ at least 40--50 MHz. The phasemeter prototype developed in Albert Einstein Institute Hannover for ESA uses 80 MHz \cite{Gerberding13}. Due to the limited bandwidth of the down link to Earth, measurement data at this high sampling rate cannot be transferred to the Earth. Instead, they are low-pass filtered and then down-sampled to a few Hz (e.g. $3\,$Hz). The raw data received on the Earth are at this sampling rate. For simulation concerns, generating measurement data at 80 MHz with a total observation time of a few years is computationally expensive and unnecessary. Instead, these measurements are directly simulated at the down-sampling rate.

It is worth clarifying that, up to this point, the simulation of the S/C and GWs was done with complete knowledge of ``mother nature"\footnote{Effectively, the ``mother nature" is the dynamic models and the noise models that we have chosen in the simulation. In the end, the outputs of the pre-processing algorithms will be compared with the true values determined by these models, hence testing the performance of the designed algorithms. }. From the next subsection, pre-data processing on, comes the simulated processing of the down-linked data, where we have only the raw data received on Earth, but other information such as the S/C status is unknown.

The next step is the so-called pre data processing \cite{Wang10}. The main task is to synchronize the raw data received at the Earth station and to determine the armlength accurately. In addition, pre data processing aims to establish a convenient framework to monitor the system performance, to compensate unexpected noise and to deal with unexpected cases such as when one laser link is broken for a short time \cite{Wang14b}. The armlength information is contained in the ranging measurements, that compare the laser transmission time at the remote S/C and the reception time at the local S/C. Since these two times are measured by different clocks, i.e. ultrastable oscillators (USOs), which have different unknown jitter and biases, the ranging data actually contain large biases. For instance, high-performance (not necessarily the best) space-qualified crystal oscillators, such as oven controlled crystal oscillators \cite{OCXO}, have a frequency stability of about $10^{-7\sim-8}$. This would lead to clock biases larger than one second in three years, which would result in huge biases in the ranging measurements. In fact, all the measurements taken in one S/C are labeled with the clock time in that S/C. This means all the time series contain clock noise. Time series from different S/C contain different clock noise. These unsynchronized, dirty and noisy time series need to be pre-processed in order to become usable for TDI.

The last two steps are after the pre data processing stage, so the pre-processing algorithms do not rely on the performance of these two steps. In the TDI step, one needs to construct TDI variables to reduce the otherwise overwhelming laser frequency noise~\cite{TDI03,TDI99,TDI03b,TDI04,TDI05,TDI02,TDI05b,TDI12}.
In the last step, the task is to dig out GW signals from the TDI variables and extract astrophysical information --- in short, detection and parameter estimation. At this stage, we have relatively clean and synchronized data labeled with UTC time stamps. Still, the GW signals are weak compared to the remaining noise. As a result, one needs to implement matched filtering techniques to obtain optimal signal-to-noise ratio (SNR) \cite{Babak08}.

\section{The interspacecraft measurements}
\label{sec:Meas1}

\begin{figure*}
\centering
\includegraphics[width=\textwidth]{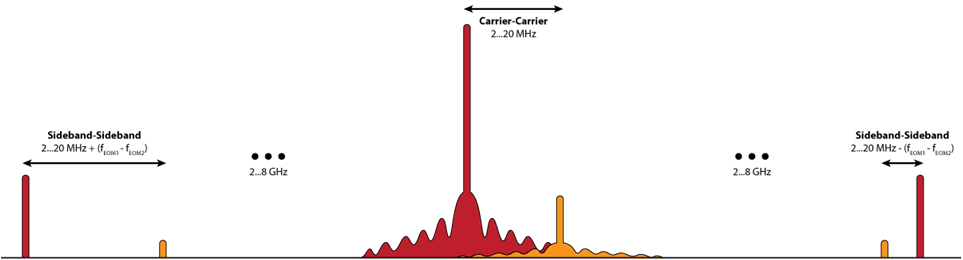}
\caption{ \label{fig:CarrierSideband} Schematic power spectral density plot of LISA carrier laser beam, clock-sideband modulation and the PRN modulation.
Horizontal direction denotes frequency and vertical direction denotes power. In the middle, the two peaks are the two beating carriers.
Around the carriers are the PRN modulations. On the sides of the figure are the clock sidebands modulation.}
\end{figure*}

Now, let us look into these inter-spacecraft measurements \cite{Barke10,Gerhard11}. In the middle of Fig.~\ref{fig:CarrierSideband}, the two peaks are the local carrier
and the weak received carrier. They form a carrier-to-carrier beatnote, which is usually called the science measurement, denoted by $f_\textrm{sci}$,
\begin{eqnarray}\label{eq:sci1}
f_\textrm{sci} = f_\textrm{Doppler} + f_\textrm{GW} + f_\textrm{noise},
\end{eqnarray}
where $f_\textrm{Doppler}$ is the Doppler shift, $f_\textrm{GW}$ is the frequency fluctuation induced by GWs, $f_\textrm{noise}$ is the noise term, which
contains various kinds of noise, such as laser frequency noise, optical path-length noise, clock noise, etc. Due to the orbits, $f_\textrm{Doppler}$
can be as large as $15\,$MHz. However, $f_\textrm{GW}$ is usually at the $\mu$Hz level. Among the noise terms, the laser frequency noise is the dominating one. The free-running laser frequency noise is expected to be above MHz/Hz${}^{1/2}$ at about $10\,$mHz. After pre-stabilization, the laser frequency noise is somewhere between $30-1000\,$Hz/Hz${}^{1/2}$ at about $10\,$mHz~\cite{WP09,Gerhard11}.

On the two sides of Fig.~\ref{fig:CarrierSideband} are the two clock sidebands. The clock sideband beatnote is given by the following
\begin{eqnarray}\label{eq:SB1}
f_\textrm{sidebandBN} = f_\textrm{Doppler} + f_\textrm{GW} + f_\textrm{noise} + m\Delta f_\textrm{clock} ,
\end{eqnarray}
where $\Delta f_\textrm{clock}$ is the frequency difference between the local USO and the remote USO, $m$ is an up-conversion factor.
Except for the intentionally amplified clock term, the clock sideband beatnote contains the same information as the carrier-to-carrier beatnote does.

The pseudo-random noise (PRN) modulations \cite{Barke10,Gerhard11,Esteban09,Esteban10,Sutton10} are around the carriers in Fig.~\ref{fig:CarrierSideband}.
The two PRN modulations shown in the figure in yellow and in red are orthogonal to each other such that no correlation
exists for any delay time. At the local S/C, one correlates the PRN code modulated on the remote laser beam with an exact
copy, hence obtaining the delay time between the emission and the reception. This light travel time tells us
the arm length information. However, the PRN codes are labelled by their own clocks at the transmitter and the receiver, respectively. Thus, the ranging signal $\tau_\textrm{ranging}$ also contains the
time difference of the two clocks.
\begin{eqnarray}\label{eq:ranging1}
\tau_\textrm{ranging} = L/c + \Delta T_\textrm{clock} + T_\textrm{noise},
\end{eqnarray}
where $L$ is the arm length, $c$ is the speed of light, $\Delta T_\textrm{clock}$ is the clock time difference,
$T_\textrm{noise}$ denotes the noise in this measurement. The ranging measurement noise $T_\textrm{noise}$ is around $3\,$ns (or $1\,$m) rms \cite{Gerhard11}.
However, since the clock is freely drifting all the time, after one year $\Delta T_\textrm{clock}$ could be
quite large. This term makes the knowledge of the armlengths much poorer than the ranging measurement noise $1\,$m rms, hence violating the requirement of TDI.  Therefore, one needs to decouple this term from the true armlength term to a level better than $3\,$ns.

\section{Formulation of the measurements}
\label{sec:Meas2}

In this section, we try to formulate the exact expressions of Eqs.~\ref{eq:sci1}, \ref{eq:SB1} and \ref{eq:ranging1}. Let us first
clarify the notation.
The positions of the S/C are denoted by $\vec{x}_i=(x_i,y_i,z_i)^T$, their velocities are denoted by $\vec{v}_i=(v_{xi},v_{yi},v_{zi})^T$
in the solar system barycenter (SSB) frame, where $i=1,2,3$ is the S/C index. Each S/C has its own USO. The measurements taken on each S/C are recorded according to their own USO. Let us denote the nominal frequency of the USO in the $i$-th S/C as $f^\textrm{nom}_i$ (the design frequency) and denote its actual frequency (the true frequency it runs at) as $f_i$. The difference
\begin{eqnarray}
\delta f_i = f_i - f^\textrm{nom}_i
\end{eqnarray}
is the frequency error of each USO. The USOs are thought to be operating at $f^\textrm{nom}_i$. The actual frequencies $f_i$ are unknown to us.
Also, we denote the nominal time of each USO as $T^\textrm{nom}_i$ (the readout time of the clock) and the actual clock time (the true time at which the clock reads
$T^\textrm{nom}_i$) as $T_i$. We have
\begin{eqnarray}
T^\textrm{nom}_i &=& \frac{\phi_i}{2\pi f^\textrm{nom}_i} = \frac{\int f_i(t) \D t}{ f^\textrm{nom}_i}, \\
T_i &=& \int \D t, \\
\phi_i &=& 2 \pi \int f_i(t) \D t,
\end{eqnarray}
where $\phi_i$ denotes the readout phase in the $i$-th S/C. The time difference
\begin{eqnarray}
\delta T_i &=& T^\textrm{nom}_i - T_i, \nonumber\\
&=& \frac{1}{f^\textrm{nom}_i}\int (f_i - f^\textrm{nom}_i) \D t, \nonumber \\
&=&\frac{1}{f^\textrm{nom}_i}\int \delta f_i \D t
\end{eqnarray}
is the clock jitter of each USO. This leads to
\begin{eqnarray}
\dot{\delta T_i} = \frac{\delta f_i}{f^\textrm{nom}_i}.
\end{eqnarray}
The above two equations mean that the clock jitter (or time jitter) is the accumulative effect of frequency jitters.
For the convenience of numerical simulations, we write the discrete version of the above formulas as follows
\begin{eqnarray}
\delta T_i(k)&=&\frac{1}{f^\textrm{nom}_i}\sum_{a=1}^k \delta f_i(a) \Delta t_s + \delta T_i(0), \\
\dot{\delta T_i}(k)&=& \frac{\delta T_i(k) - \delta T_i(k-1)}{\Delta t_s} \nonumber \\
&=& \frac{\delta f_i(k) \Delta t_s / f^\textrm{nom}_i}{\Delta t_s} \nonumber \\
&=& \frac{\delta f_i(k)}{f^\textrm{nom}_i},
\end{eqnarray}
where $k$ in the parentheses means the value at the $k$-th step or at time $k\Delta t_s$, $\delta T_i(0)$
stands for the initial clock bias.

To this point, we try to formulate the ranging measurements. For convenience, we write it in dimensions of length and denote the armlength measurements
measured by the laser link from S/C $i$ to S/C $j$ (measured at S/C $j$) as $R_{ij}$. Thus, we have
\begin{eqnarray}\label{eq:ranging2}
R_{ij}(k) = L_{ij}(k) + [ \delta T_j(k) - \delta T_i(k) ]c + \textrm{noise},
\end{eqnarray}
where $L_{ij}(k)$ is the true armlength we want to obtain from the ranging measurements, $[ \delta T_j(k) - \delta T_i(k) ]c$ is
the armlength bias caused by the clock jitter, and ``noise" denotes the effects of other noise sources. Notice that
the step $k$ corresponds to the uniform recording time $k \Delta T_s$, which means the clock errors are not included in the recording
time yet, but only in the measurements. Also notice that $\delta T_i(k)$ is the clock error of the remote S/C $i$ at the current time.
This is a second approximation we have made in this paper, since the delay $R_{ij}/c$ is only simulated as the measurements, but not
in the recording time. (See more discussions in the summary.)

Next, we want to consider Doppler measurements or science measurements. They are phase measurements recorded at the phasemeter. For convenience,
we formulate them as frequency measurements, since it is trivial to convert phase measurements to frequency measurements. First, we take into
account only the imperfection of the USO and ignore other noises. We denote the true frequency we want to measure as $f_\textrm{true}$ and the frequency
actually measured as $f_\textrm{meas}$. The USO is thought to be running at $f^\textrm{nom}$. The recorded frequency $f_\textrm{meas}$ is compared
to it. However, the frequency at which the USO is really running is $f=f^\textrm{nom} + \delta f$. This is what the true frequency $f_\textrm{true}$ is
actually compared to. Thus, we have the following formula
\begin{eqnarray}
\frac{f_\textrm{meas}}{f^\textrm{nom}} &=& \frac{f_\textrm{true}}{f} \nonumber \\
&=& \frac{f_\textrm{true}}{f^\textrm{nom} + \delta f}.
\end{eqnarray}
For a normal USO, $\delta f / f^\textrm{nom}$ is usually a very small number ($<10^{-8}$), therefore the second order in it is smaller than machine
accuracy. Thus, we can write the above equation in linear order of $\delta f / f^\textrm{nom}$ for numerical simulation concern without loss of
precision:
\begin{eqnarray}
f_\textrm{meas} &=& \frac{f_\textrm{true}}{1 + \delta f / f^\textrm{nom}} \nonumber \\
&=& f_\textrm{true}\left(1 - \frac{\delta f}{f^\textrm{nom}}\right).
\end{eqnarray}
We denote the average carrier frequency (the average laser frequency over certain time) as $f^\textrm{carrier}$, the laser frequency noise as $\delta f^\textrm{c}$ and the unit vector pointing from S/C $i$ to S/C $j$ as $\hat{n}_{ij}$. Let us consider the laser link sent from S/C $i$ to S/C $j$. When transmitted at S/C $i$, the instantaneous carrier frequency is actually $f_i^\textrm{carrier}+\delta f_i^\textrm{c}$. When received at S/C $j$, this carrier frequency has been Doppler shifted and the GW signals are encoded. Therefore, the received carrier frequency at S/C $j$ can be written as
\begin{eqnarray}
(f_i^\textrm{carrier}+\delta f_i^\textrm{c})\left[ 1 - \frac{(\vec{v}_j-\vec{v}_i)\cdot \hat{n}_{ij}}{c} \right] - f^\textrm{GW}_{ij}.
\end{eqnarray}
This carrier is then beat with the local carrier $f_j^\textrm{carrier}+\delta f_j^\textrm{c}$ of S/C $j$. The resulting beatnote is
the science measurement
\begin{widetext}
\begin{eqnarray}\label{eq:sci2}
f^\textrm{sci}_{ij}(k) &=& \left[  f^\textrm{carrier}_j - f^\textrm{carrier}_i \left( 1 - \frac{(\vec{v}_j-\vec{v}_i)\cdot \hat{n}_{ij}}{c} \right) + f^\textrm{GW}_{ij}(k)\right]\left( 1- \frac{\delta f_j(k)}{f^\textrm{nom}_j}\right) \nonumber  \\
&& + \left[ \delta f^\textrm{c}_j - \delta f^\textrm{c}_i \left( 1 - \frac{(\vec{v}_j-\vec{v}_i)\cdot \hat{n}_{ij}}{c} \right) \right]\left( 1- \frac{\delta f_j(k)}{f^\textrm{nom}_j}\right) + \textrm{noise}, \nonumber \\
&=&\left[  f^\textrm{carrier}_j - f^\textrm{carrier}_i \left( 1 - \frac{(\vec{v}_j-\vec{v}_i)\cdot \hat{n}_{ij}}{c} \right) + f^\textrm{GW}_{ij}(k)\right]\left( 1- \frac{\delta f_j(k)}{f^\textrm{nom}_j}\right)+\textrm{noise}, \nonumber \\
\end{eqnarray}
\end{widetext}
where in the last step we have absorbed the laser frequency noise into the noise term.
In practice, the carrier frequencies are adjusted occasionally (controlled by a pre-determined frequency plan) to make sure that the carrier-to-carrier beatnote is within a certain frequency range. Hence, $f^\textrm{carrier}_i$ is also a function of time.

Now, let us consider the clock sidebands. At S/C $i$, the clock frequency $f_i^\textrm{nom}+\delta f_i$ is up-converted by a factor $m_i$, which is about $40-50$, and modulated onto
the carrier through an electro optical modulator (EOM). Therefore, we have an upper clock sideband and a lower clock sideband as follows
\begin{eqnarray}
f_i^\textrm{USB} &=& f_i^\textrm{carrier}+\delta f_i^\textrm{c} + m_i (f_i^\textrm{nom}+\delta f_i), \\
f_i^\textrm{LSB} &=& f_i^\textrm{carrier}+\delta f_i^\textrm{c} - m_i (f_i^\textrm{nom}+\delta f_i).
\end{eqnarray}
When received by S/C $j$, both the Doppler effect and GWs are present. Therefore, the received frequencies (at S/C $j$) of the upper and the lower
clock sideband are as follows
\begin{eqnarray}
&&\left[f_i^\textrm{carrier}+\delta f_i^\textrm{c} \pm m_i (f_i^\textrm{nom}+\delta f_i)\right] \nonumber \\
&&\cdot \left[ 1 - \frac{(\vec{v}_j-\vec{v}_i)\cdot \hat{n}_{ij}}{c} \right] - f^\textrm{GW}_{ij}.
\end{eqnarray}
The clock sideband beatnote is obtained by beating this frequency with the local clock sideband
\begin{widetext}
\begin{eqnarray}
f^\textrm{sidebandBN}_{ij}(k) &=& \left[  f^\textrm{carrier}_j - f^\textrm{carrier}_i \left( 1 - \frac{(\vec{v}_j-\vec{v}_i)\cdot \hat{n}_{ij}}{c} \right) + f^\textrm{GW}_{ij}(k)\right]\left( 1- \frac{\delta f_j(k)}{f^\textrm{nom}_j}\right)\nonumber \\
&+& \left[  m_j (f_j^\textrm{nom}+\delta f_j(k)) - m_i (f_i^\textrm{nom}+\delta f_i(k)) \left( 1 - \frac{(\vec{v}_j-\vec{v}_i)\cdot \hat{n}_{ij}}{c} \right) \right] \nonumber \\
&& \cdot \left( 1- \frac{\delta f_j(k)}{f^\textrm{nom}_j}\right) + \textrm{noise}, \nonumber \\
&=& \left[  f^\textrm{carrier}_j - f^\textrm{carrier}_i \left( 1 - \frac{(\vec{v}_j-\vec{v}_i)\cdot \hat{n}_{ij}}{c} \right) + f^\textrm{GW}_{ij}(k)\right]\left( 1- \frac{\delta f_j(k)}{f^\textrm{nom}_j}\right)\nonumber \\
&+& [\alpha_j \delta f_j(k) - \alpha_i \delta f_i(k)] + ( m_j f_j^\textrm{nom} - m_i f_i^\textrm{nom} ) + m_i f_i^\textrm{nom} \frac{(\vec{v}_j-\vec{v}_i)\cdot \hat{n}_{ij}}{c}, \nonumber \\
&+& \textrm{noise}
\end{eqnarray}
\end{widetext}
where $\alpha_i$ and $\alpha_j$ are some known constants. Notice that we have neglected some minor terms in the last step.
For simulation purposes, we temporarily ignore the constant term $m_j f_j^\textrm{nom} - m_i f_i^\textrm{nom}$ and the small Doppler term $m_i f_i^\textrm{nom} (\vec{v}_j-\vec{v}_i)\cdot \hat{n}_{ij}/c$. Furthermore, we write $\alpha_i$ and $\alpha_j$ as a uniform up-conversion factor $m$ for simplicity.
Then, we have the simplified formula
\begin{widetext}
\begin{eqnarray}\label{eq:SB2}
f^\textrm{sidebandBN}_{ij}(k)&=&\left[  f^\textrm{carrier}_j - f^\textrm{carrier}_i \left( 1 - \frac{(\vec{v}_j-\vec{v}_i)\cdot \hat{n}_{ij}}{c} \right) + f^\textrm{GW}_{ij}(k)\right]\left( 1- \frac{\delta f_j(k)}{f^\textrm{nom}_j}\right)\nonumber \\
&+&   m(\delta f_j(k) - \delta f_i(k))   + \textrm{noise}.
\end{eqnarray}
\end{widetext}
Up to now, we have formulated all the inter-spacecraft measurements in Eqs.~\ref{eq:ranging2}, \ref{eq:sci2} and \ref{eq:SB2}.

\section{The hybrid extended Kalman filter}

\label{sec:Kalman}

The hybrid extended Kalman filter\cite{Dan} is designed for a system with continuous and nonlinear dynamic equations along with nonlinear measurement
equations. First, we describe the model of such systems as follows
\begin{eqnarray}
\dot{x} &=& f(x,t) + w(t)  \label{eq:fx}  \\
y_k &=& h_k(x_k, v_k) \label{eq:hx} \\
E[w(t)w^T(t+\tau)] &=& W_c \delta (\tau) \label{eq:wc}\\
v_k &\sim& (0,V_k),  \label{eq:vk}
\end{eqnarray}
where both the dynamic function $f(x,t)$ and the measurement function $h_k(x_k)$ are nonlinear, $w(t)$ is the
continuous noise.  $x,f(x,t),w,y_k,h_k(x_k),v_k$ are column vectors. $W_c,W_k$ are covariance matrices. If we discretize the noise with a step size $\Delta t$, we have
\begin{eqnarray}
w_k \sim (0,W_k), \label{eq:wk}
\end{eqnarray}
where it can be proven that $W_k = W_c(k\Delta t)/{\Delta t}$. In order to fit Eqs.~\ref{eq:fx}, \ref{eq:hx}, \ref{eq:wc}, \ref{eq:vk} into
the standard Kalman filter frame, we need to linearize and discretize the formulae and solve the dynamic equation.
Eq.~\ref{eq:fx} is expanded to linear order in $x_0$ as follows
\begin{eqnarray}\label{eq:fx_linear}
\dot{x} &\approx& f(x_0,t_0) + \left.\frac{\partial f}{\partial x}\right|_{x_0,t_0}(x-x_0) + w(t) \nonumber \\
&=& f(x_0,t_0) + F(x_0,t_0)(x-x_0) + w(t),
\end{eqnarray}
where we have defined $F(x_0,t_0)\equiv\left.\frac{\partial f}{\partial x}\right|_{x_0,t_0}$, and assumed $\frac{\partial f}{\partial t} \ll 1$. The expectation of this linearized equation (where $\textrm{E}[w(t)]=0$ is used) can be solved exactly as follows
\begin{eqnarray}\label{eq:x_dyn}
x(t_2) &=& e^{F(x_0,t_0)\Delta t} x(t_1)   \nonumber \\
       &+& \left[e^{F(x_0,t_0)\Delta t}-I\right]\left[F^{-1}(x_0,t_0) f(x_0,t_0) - x_0 \right], \nonumber \\
\end{eqnarray}
where $\Delta t = t_2 - t_1$, and the matrix exponential is defined as
\begin{eqnarray}
e^{F\Delta t} \equiv \sum_{n=0}^{+\infty} \frac{(F\Delta t)^n}{n!}.
\end{eqnarray}
Now, let us switch to the standard Kalman filter notation and denote $x(t_2),x(t_1)$ and $F(x_0,t_0)$ as $\hat{x}_k^-,\hat{x}^+_{k-1}$ and $F_{k-1}$, respectively.
Eq.~\ref{eq:x_dyn} can be rewritten as
\begin{eqnarray}
\hat{x}_k^- &=& e^{F_{k-1}\Delta t} \hat{x}^+_{k-1}  \nonumber \\
            &+& (e^{F_{k-1}\Delta t}-I)\left[F_{k-1}^{-1} f(x_0,t_0) - x_0 \right].
\end{eqnarray}
Notice that $x_0$ is a nominal trajectory, around which the Taylor expansion is made.
Based on the above solution, the propagation equation of the covariance matrices is obtained
\begin{eqnarray}
P_k^- = e^{F_{k-1}\Delta t} P^+_{k-1} e^{F^T_{k-1}\Delta t} + W_{k-1},
\end{eqnarray}
where $P^-,P^+$ are the a priori and a posteriori covariance matrices as before. Alternatively, Eq.~\ref{eq:fx_linear} can be solved
approximately by converting the differential equation to a difference equation. The corresponding formulae are
\begin{eqnarray}
\hat{x}_k^- &=& (I + F_{k-1}\Delta t) \hat{x}^+_{k-1} + \left[f(x_0,t_0)-F_{k-1} x_0 \right] \Delta t, \\
P_k^- &=& (I + F_{k-1}\Delta t) P^+_{k-1} (I + F_{k-1}\Delta t)^T + W_{k-1}.
\end{eqnarray}
The above two equations can also be obtained from the exact solutions by replacing $e^{F_{k-1}\Delta t}$ with $I+F_{k-1}\Delta t$.
The advantage of these formulae is that they are computationally less expensive. On the other hand, they are less precise. The measurement formula can be
linearized similarly
\begin{eqnarray}
y_k = H_k x_k + [h_k(\hat{x}^-_k,0)-H_k \hat{x}^-_k] + M_k v_k,
\end{eqnarray}
where $H_k\equiv \left.\frac{\partial h_k}{\partial x}\right|_{\hat{x}^-_k}, M_k\equiv \left.\frac{\partial h_k}{\partial v}\right|_{\hat{x}^-_k}$.
Now, the Kalman filter can be applied without much effort. We summarize the hybrid extended Kalman filter formulae for the model described by
Eqs.~\ref{eq:fx}, \ref{eq:hx}, \ref{eq:wc}, \ref{eq:vk} as follows:
\begin{enumerate}
  \item Initialize the state vector and the covariance matrix
  \begin{eqnarray}
  \hat{x}_0^+, P_0^+.
  \end{eqnarray}
  \item Calculate the a priori estimate $\hat{x}_k^-$ from the a posteriori estimate $\hat{x}^+_{k-1}$ at the previous step, using the dynamic equation
  \begin{eqnarray}
  \dot{x} &=& f(x,t).
  \end{eqnarray}
  Use either of the following two formulae to update the covariance matrix
   \begin{eqnarray}
   P_k^- &=& e^{F_{k-1}\Delta t} P^+_{k-1} e^{F^T_{k-1}\Delta t} + W_{k-1}, \\
   P_k^- &=& (I + F_{k-1}\Delta t) P^+_{k-1} (I + F_{k-1}\Delta t)^T + W_{k-1}.
   \end{eqnarray}
  \item Calculate the Kalman gain
  \begin{eqnarray}
  K_k = P^-_k H^T_k (H_k P_k^- H_k^T + M_k V_k M_k^T)^{-1}.
  \end{eqnarray}
  \item Correct the a priori estimate
  \begin{eqnarray}
  \hat{x}_k^+ &=& \hat{x}^-_k + K_k [y_k - h_k(\hat{x}^-_k,0)], \\
  P^+_k &=& (I - K_k H_k)P_k^-,    \nonumber \\
  &=& (I - K_k H_k) P^-_k (I - K_k H_k)^T + K_k V_k K_k^T.
  \end{eqnarray}
\end{enumerate}

\section{Kalman filter model for LISA}

\label{sec:LISAmodel}

In this section, we want to design a hybrid extended Kalman filter for LISA. First, we define
a 24-dimensional column state vector
\begin{eqnarray}
x=(\vec{x}_1, \vec{x}_2, \vec{x}_3, \vec{v}_1, \vec{v}_2, \vec{v}_3, \delta T_1, \delta T_2, \delta T_3,
\delta f_1, \delta f_2, \delta f_3)^T,   \nonumber
\end{eqnarray}
where $\vec{x}_i=(x_i,y_i,z_i)^T$ are the S/C positions, $\vec{v}_i=(v_{xi},v_{yi},v_{zi})^T$ are the S/C velocities,
$\delta T_i$ and $\delta f_i$ are the clock jitters and frequency jitters, and $i=1,2,3$ is the S/C index. Please note
the difference between the state vector $x_k$, the measurements $y_k$, and the
position components $(x_i,y_i,z_i)$, since the latter index is the S/C label and can only take three values $1,2,3$. For convenience,
we rewrite the measurement formulae derived . The ranging measurements from S/C $i$ to S/C $j$ are
\begin{eqnarray}\label{eq:R}
R_{ij} &=& L_{ij} + (\delta T_j - \delta T_i) c + n^R_{ij}     \nonumber \\
&=&  \sqrt{(x_j-x_i)^2+(y_j-y_i)^2+(z_j-z_i)^2}  \nonumber \\
&+& (\delta T_j - \delta T_i)\cdot c + n^R_{ij},
\end{eqnarray}
where $n^R_{ij}$ is the ranging measurement noise. The Doppler measurements are denoted as $D_{ij}$,
\begin{eqnarray}\label{eq:D}
D_{ij} &=& \left[  f^\textrm{carrier}_j - f^\textrm{carrier}_i \left( 1 - \frac{(\vec{v}_j-\vec{v}_i)\cdot \hat{n}_{ij}}{c} \right) \right. \nonumber \\
&+& \left. f^\textrm{GW}_{ij}\right]\left( 1- \frac{\delta f_j}{f^\textrm{nom}_j}\right)+n^D_{ij},
\end{eqnarray}
where $n^D_{ij}$ is the Doppler measurement noise. Since the sideband measurements contain the same information as the Doppler measurements, in addition
the amplified differential clock jitters, we take the difference. Then, we divide both sides of the equation by the up-conversion factor $m$ and denote it as the clock measurements $C_{ij}$.
\begin{eqnarray}\label{eq:C}
C_{ij} = \delta f_j - \delta f_i + n^C_{ij},
\end{eqnarray}
where $n^C_{ij}$ is the corresponding measurement noise, and the up-conversion factor $m$ has already been absorbed into $n^C_{ij}$. Altogether, we have 18 measurement formulae, summarized in the 18-dimensional
column measurement vector
\begin{eqnarray}
y &=& h(x,v), \nonumber \\
&=& (R_{31},D_{31},C_{31},R_{21},D_{21},C_{21},R_{12},D_{12},C_{12},... \nonumber \\
      &&R_{32},D_{32},C_{32},R_{23},D_{23},C_{23},R_{13},D_{13},C_{13})^T,  \nonumber
\end{eqnarray}
where $v$ is the measurement noise. The 18-by-24 matrix $H_k$ and the 18-by-18 matrix $M_k$ can thus be calculated analytically. We omit the explicit expressions
of the 432 components in $H_k$ here. As an example, we show the $[1,1]$ component of $H_k$ omitting the step index $k$ as follows
\begin{eqnarray}
H[1,1] &=& \frac{\partial R_{31}}{\partial x_1} \nonumber \\
&=& \frac{x_1-x_3}{\sqrt{(x_1-x_3)^2+(y_1-y_3)^2+(z_1-z_3)^2}}.
\end{eqnarray}
As for $M_k$, if the dependence of the measurements $y_k$ on the noise is linear and without cross coupling, it is simply an identity matrix.

Next, we want to construct the dynamic model for the Kalman filter. Let us consider
the solar system dynamics for a single S/C. To Newtonian order the solar system dynamics can be written as
\begin{eqnarray}
\sum_i \frac{GM_i}{r_i^3}\vec{r}_i=\ddot{\vec{x}}
\end{eqnarray}
where $\vec{x}$ is the position of one LISA S/C,
$M_i,\vec{x}_i$ are the mass and the coordinates of the $i$th
celestial body (the Sun and the planets) in the solar system,
$\vec{r}_i=\vec{x}_i-\vec{x}$ is a vector pointing from that
S/C to the $i$th celestial body, $r_i=|\vec{x}_i-\vec{x}|$. The dynamic
equation can be written in a different form
\begin{eqnarray}
\frac{\mathrm{d}}{\mathrm{d}t}
\begin{bmatrix}
\vec{x} \\
\vec{v}
\end{bmatrix}
&=& f(\vec{x},\vec{v}) \nonumber \\
&=&
\begin{bmatrix}
\vec{v} \\
\sum_i GM_i (\vec{x}_i-\vec{x})/r_i^3
\end{bmatrix}.
\end{eqnarray}
We denote $\theta=(\vec{x},\vec{v})^T$, thus
\begin{eqnarray}
F=\frac{\partial f}{\partial \theta}=
\begin{bmatrix}
\mathbf{O}_3 & \mathbf{I}_3 \\
\mathbf{A}   & \mathbf{O}_3
\end{bmatrix},
\end{eqnarray}
where $\mathbf{O}_3$ denotes a 3-by-3 zero matrix, $\mathbf{I}_3$ denotes a 3-by-3 identity matrix,
and the 3-by-3 matrix $\mathbf{A}$ is defined as follows
\begin{eqnarray}
\mathbf{A} &=& - \sum_i \frac{GM_i}{r_i^3} \mathbf{I}_3 + \sum_i \frac{3GM_i}{r_i^5}(\vec{x}_i-\vec{x})(\vec{x}_i-\vec{x})^T.
\end{eqnarray}
The dynamic equation for the clock jitters and frequency jitters depends on the specific clock and how
well we characterize the clock. A simple dynamic model is shown as follows
\begin{eqnarray}
\frac{\mathrm{d}}{\mathrm{d}t}
\begin{bmatrix}
\delta T \\
\delta f
\end{bmatrix}
=
\begin{bmatrix}
\delta f / f^\mathrm{nom} \\
0
\end{bmatrix},
\end{eqnarray}
where $\delta T, \delta f$ denote clock jitters and frequency jitters. For the whole LISA constellation,
the dynamic matrix $F=\left.\frac{\partial f}{\partial x}\right|$ is 24-by-24. We omit its explicit expression
here, since it can be obtained straightforwardly from the above formulae.

\section{Simulation results}

\label{sec:Simulation}

We simulated LISA measurements of about $1400$ seconds with a sampling frequency of $3\,$Hz. Since there
are only two independent clock biases out of three, we set one clock bias to be zero, thus defining this clock as reference. The other two initial
clock biases are randomly drawn from a Gaussian distribution with a standard deviation of $0.1\,$s. This would
in turn cause a bias of about $4.2\times10^7\,$m in the ranging measurements. The (unknown) initial frequency offset of each USO
is randomly drawn from a Gaussian distribution with a standard deviation of $1\,$Hz. The frequency jitter of each USO
has a linear spectral density $ (9.2\times 10^{-6} \textrm{Hz} / f )\,$Hz$/\sqrt{\textrm{Hz}}$.
Additionally, we assume the ranging measurement noise to be white Gaussian with a standard deviation of $1\,$m.
The linear spectral density of the pre-stabilized laser is assumed to be $400\,\textrm{Hz}/\sqrt{\textrm{Hz}}$.
The clock measurement noise is white Gaussian with a standard deviation of $1\,$Hz.

\begin{figure}
\includegraphics[width=0.5\textwidth]{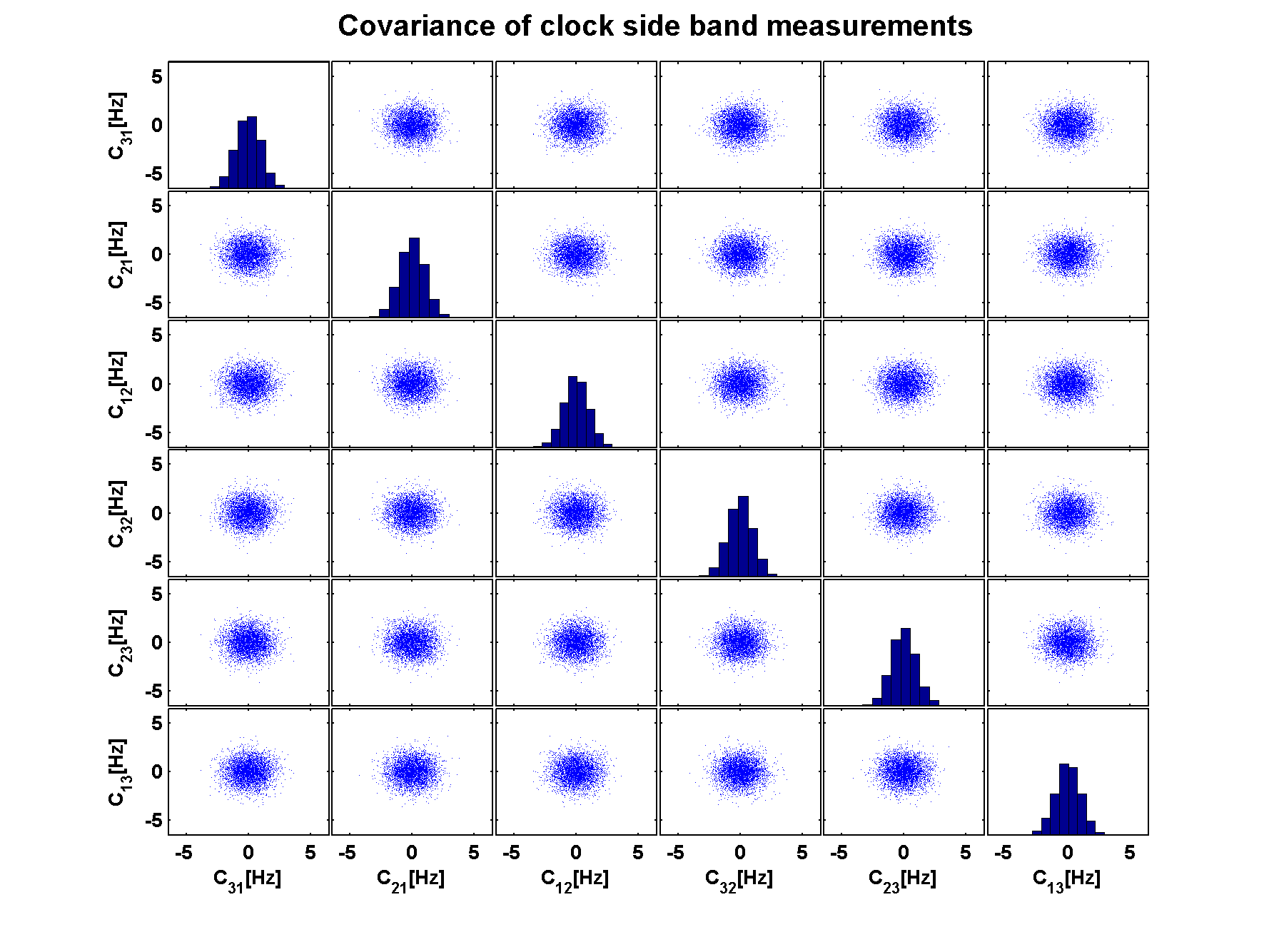}
\caption{ \label{fig:Cov_C} Scatter plot of clock measurements $C_{ij}$.
}
\end{figure}

\begin{figure}
\includegraphics[width=0.5\textwidth]{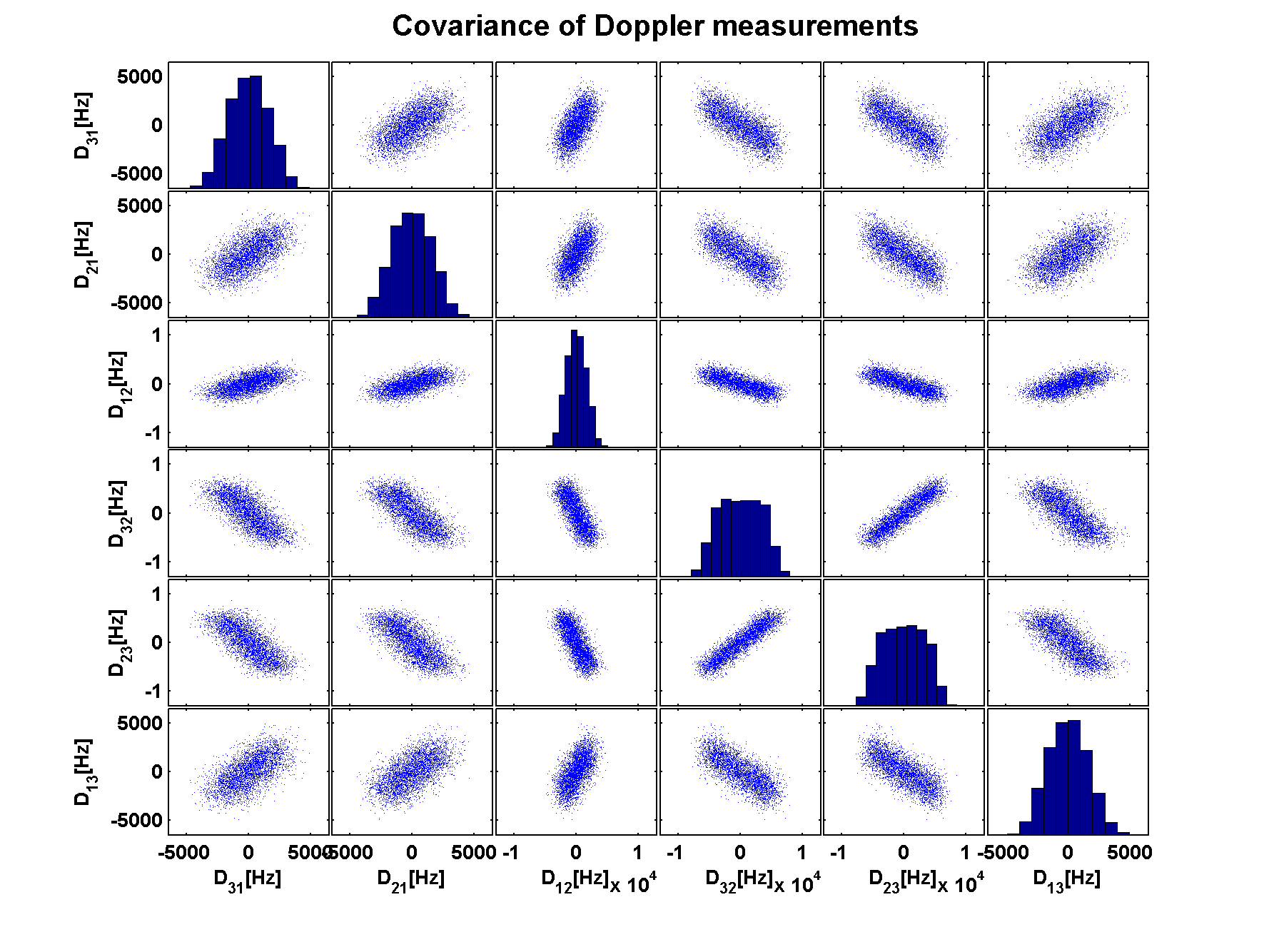}
\caption{ \label{fig:Cov_D} Scatter plot of Doppler measurements $D_{ij}$. Unlike clock measurements, scatter plots of Doppler measurements exhibit
 elliptical clouds.
 }
\end{figure}

\begin{figure}
\includegraphics[width=0.5\textwidth]{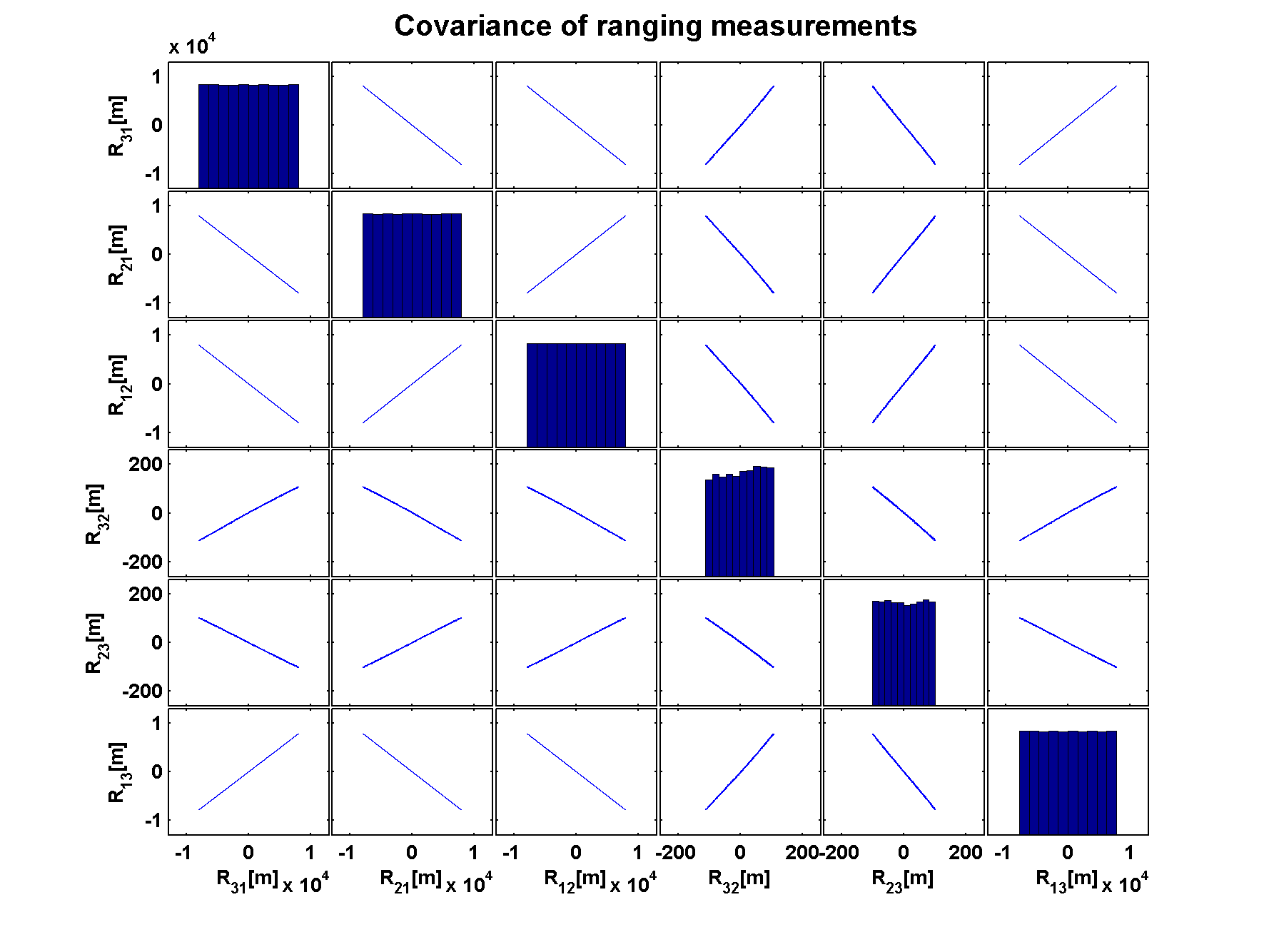}
\caption{ \label{fig:Cov_R} Scatter plot of ranging measurements $R_{ij}$. The armlength variation is much larger than the ranging measurement noise. Therefore,
we see only lines in the off-diagonal plots, which mainly show the armlength changes. The ranging measurement noise is too small compared to the armlength change to be visible in the plot.}
\end{figure}

\begin{figure}
\includegraphics[width=0.5\textwidth]{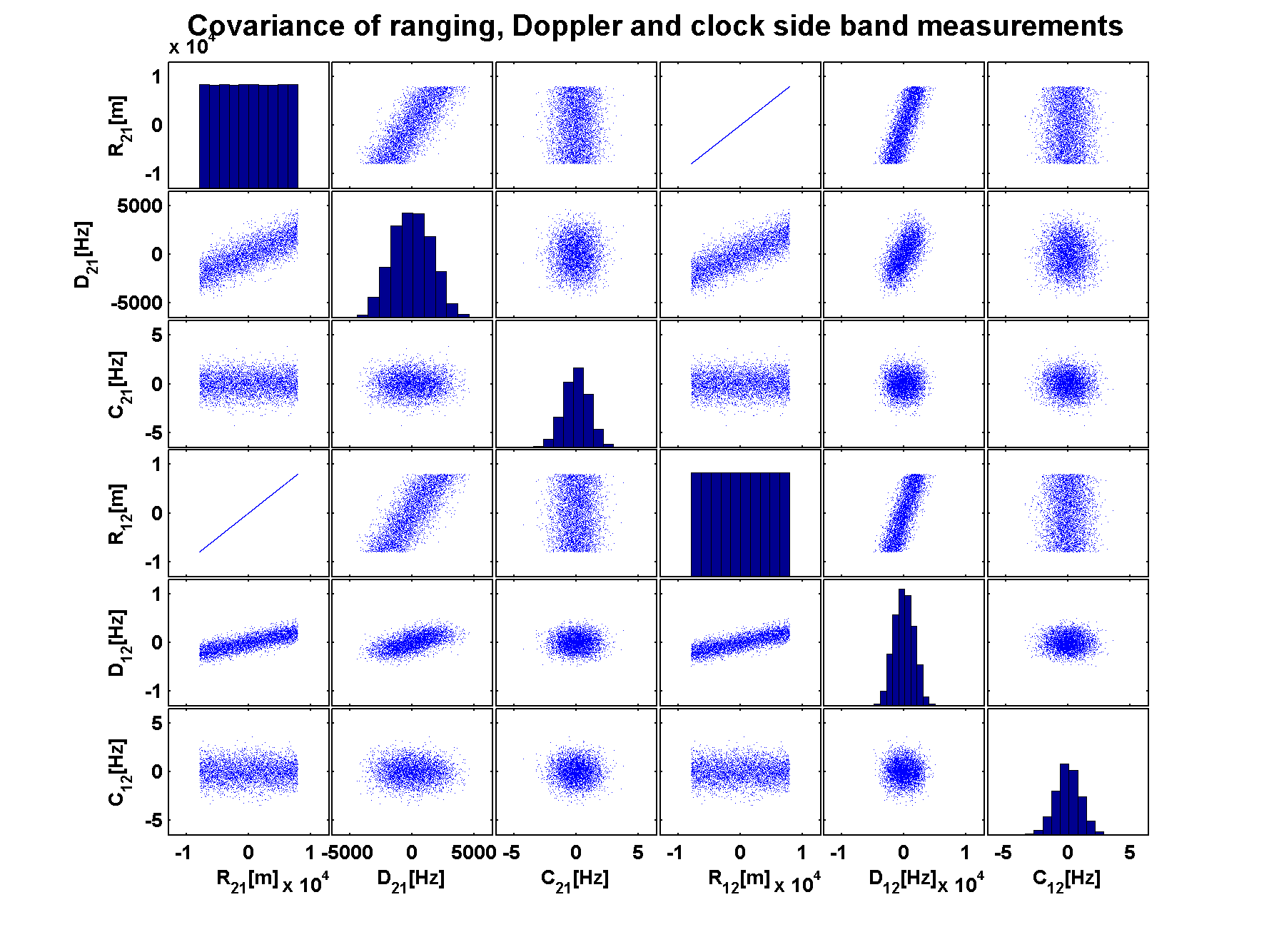}
\caption{ \label{fig:Cov_RDC} Scatter plot of different measurements $C_{ij},D_{ij},R_{ij}$. Ranging measurements are correlated with Doppler measurements, but neither
of them are correlated with clock measurements.}
\end{figure}

We show the scatter plots of the measurements $R_{ij},D_{ij},C_{ij}$ in Figs.~\ref{fig:Cov_C}, \ref{fig:Cov_D}, \ref{fig:Cov_R} and \ref{fig:Cov_RDC}. Notice that
the average of all the measurements has been removed in the plots for clarity.
Fig.~\ref{fig:Cov_C} is a scatter plot of the clock measurements $C_{ij}$. The frequency drifts within $1400\,$s are much smaller than
the clock measurement noise. Thus, they are buried in the uncorrelated clock measurement noise in the plot. The diagonal histograms show
that each clock measurement channel behaves like Gaussian noise during short observation times. The off-diagonal scatter plots are roughly circular scattering clouds, showing that different clock measurement channels are roughly uncorrelated within short times. Unlike clock measurements, scatter plots of Doppler measurements in Fig.~\ref{fig:Cov_D} exhibit elliptical clouds. This is because the Doppler shift whin $1400\,$s is sizable, which leads to the trend in the plot. The slope of the major axis of the ellipse indicates whether the two Doppler measurement channels are positively correlated or anti-correlated. The real armlength variation is much larger than the ranging measurement noise. Therefore, we see only lines in the off-diagonal plots in Fig.~\ref{fig:Cov_R}, which mainly show the armlength changes. The ranging measurement noise is too small compared to the armlength change to be visible in the plot. Fig.~\ref{fig:Cov_RDC} shows scatter plots of different measurements $C_{ij},D_{ij},R_{ij}$. It is seen from the plot that ranging measurements are correlated with Doppler measurements, but neither of them are correlated with clock measurements.

\begin{figure}[htbp]
\centering
\subfloat[$P_1^-$.]{
\begin{minipage}[t]{0.25\textwidth}
\centering
\includegraphics[width=1.0\textwidth]{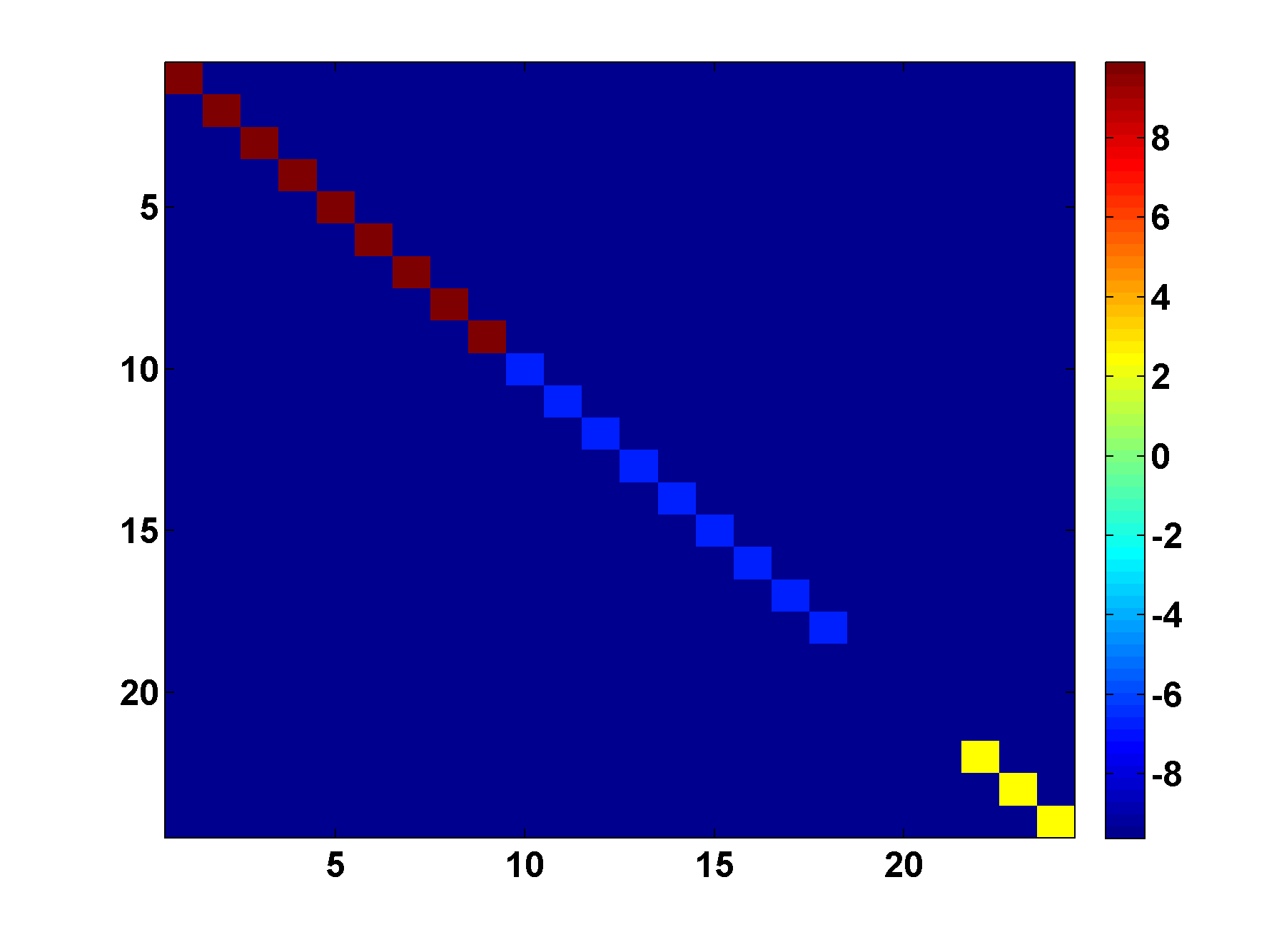}
\end{minipage}
}
\subfloat[$P_2^-$.]{
\begin{minipage}[t]{0.25\textwidth}
\centering
\includegraphics[width=1.0\textwidth]{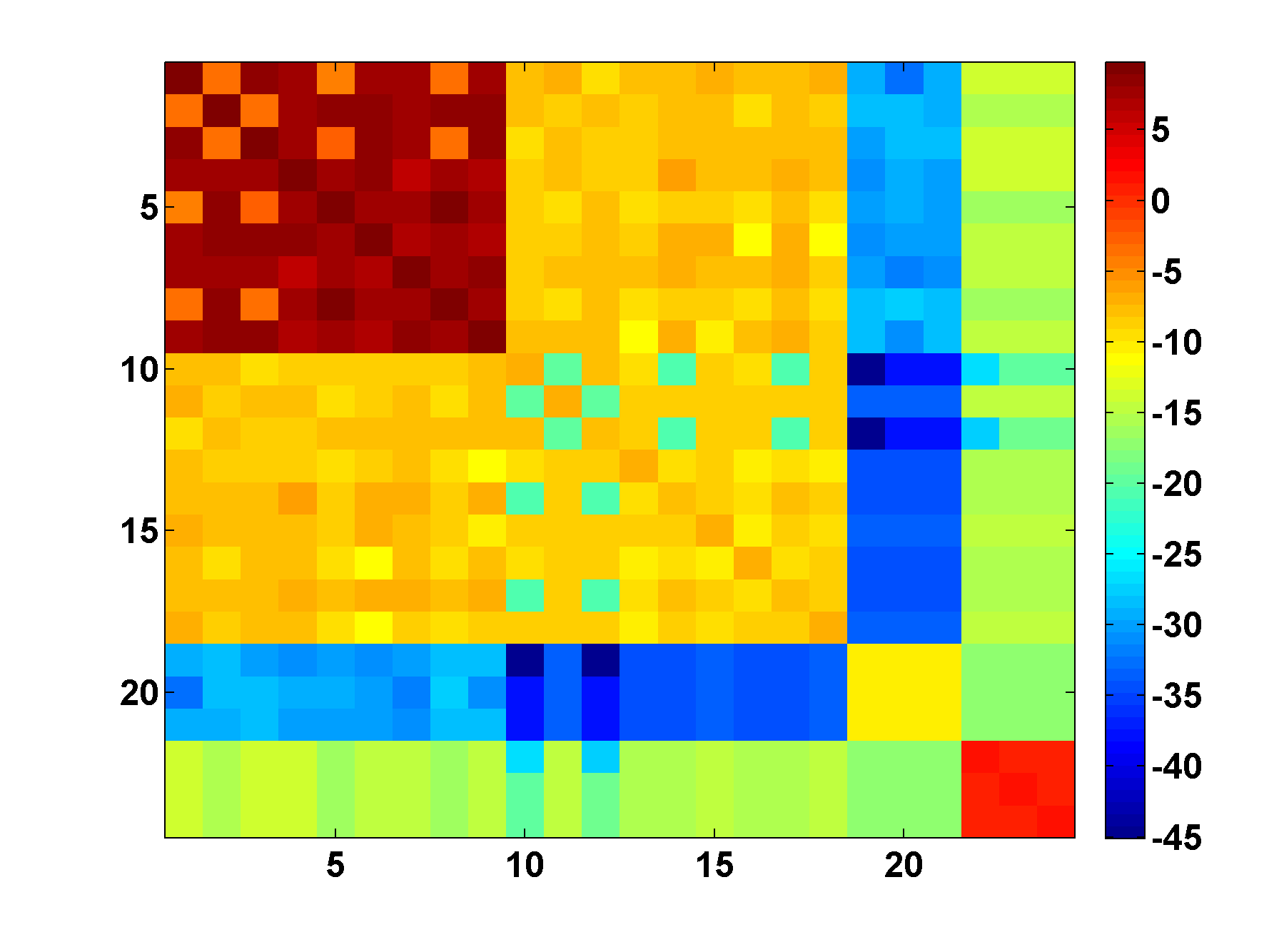}
\end{minipage}
}
\\
\subfloat[$P_5^-$.]{
\begin{minipage}[t]{0.25\textwidth}
\centering
\includegraphics[width=1.0\textwidth]{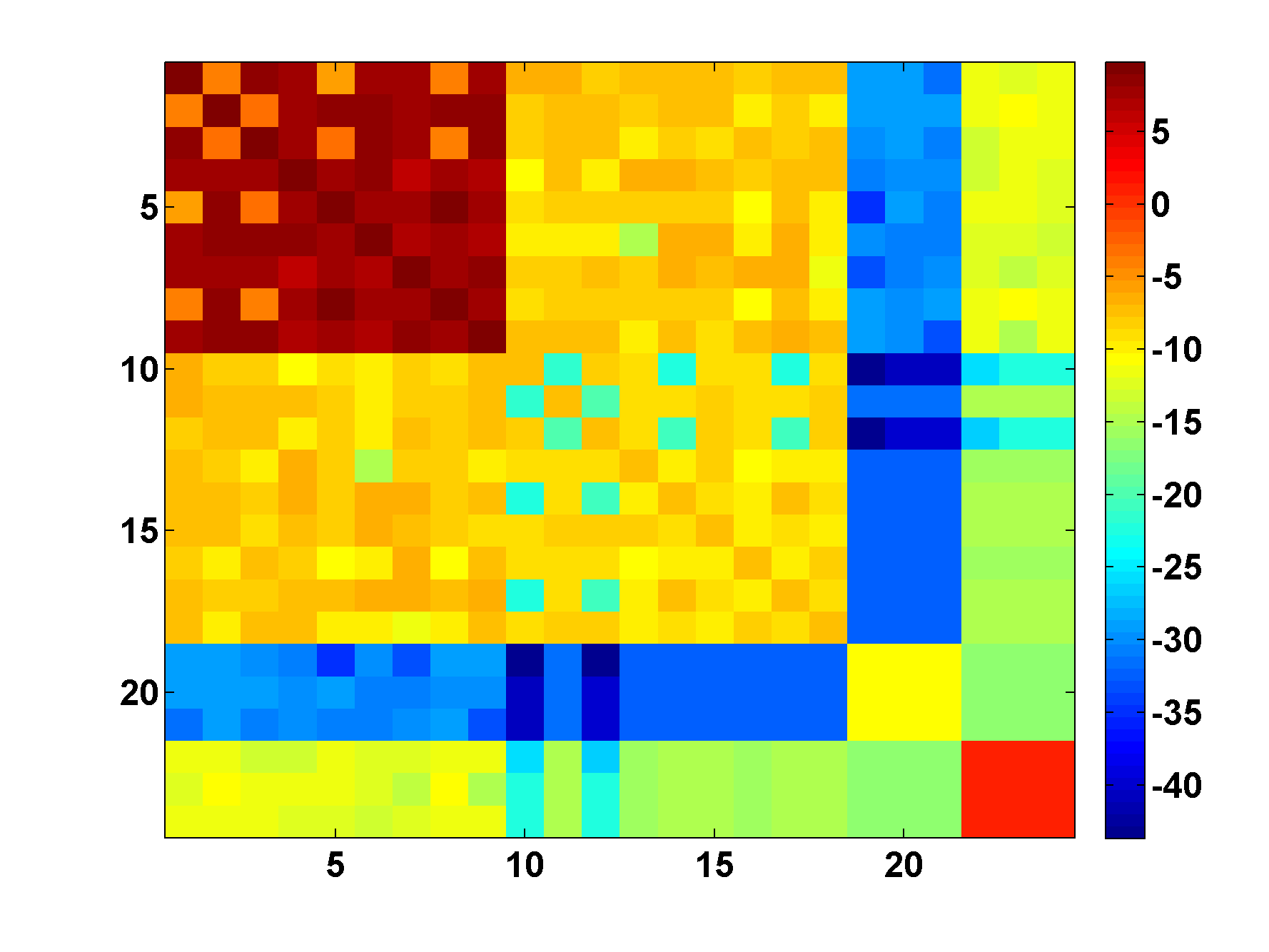}
\end{minipage}
}
\subfloat[$P_{10}^-$.]{
\begin{minipage}[t]{0.25\textwidth}
\centering
\includegraphics[width=1.0\textwidth]{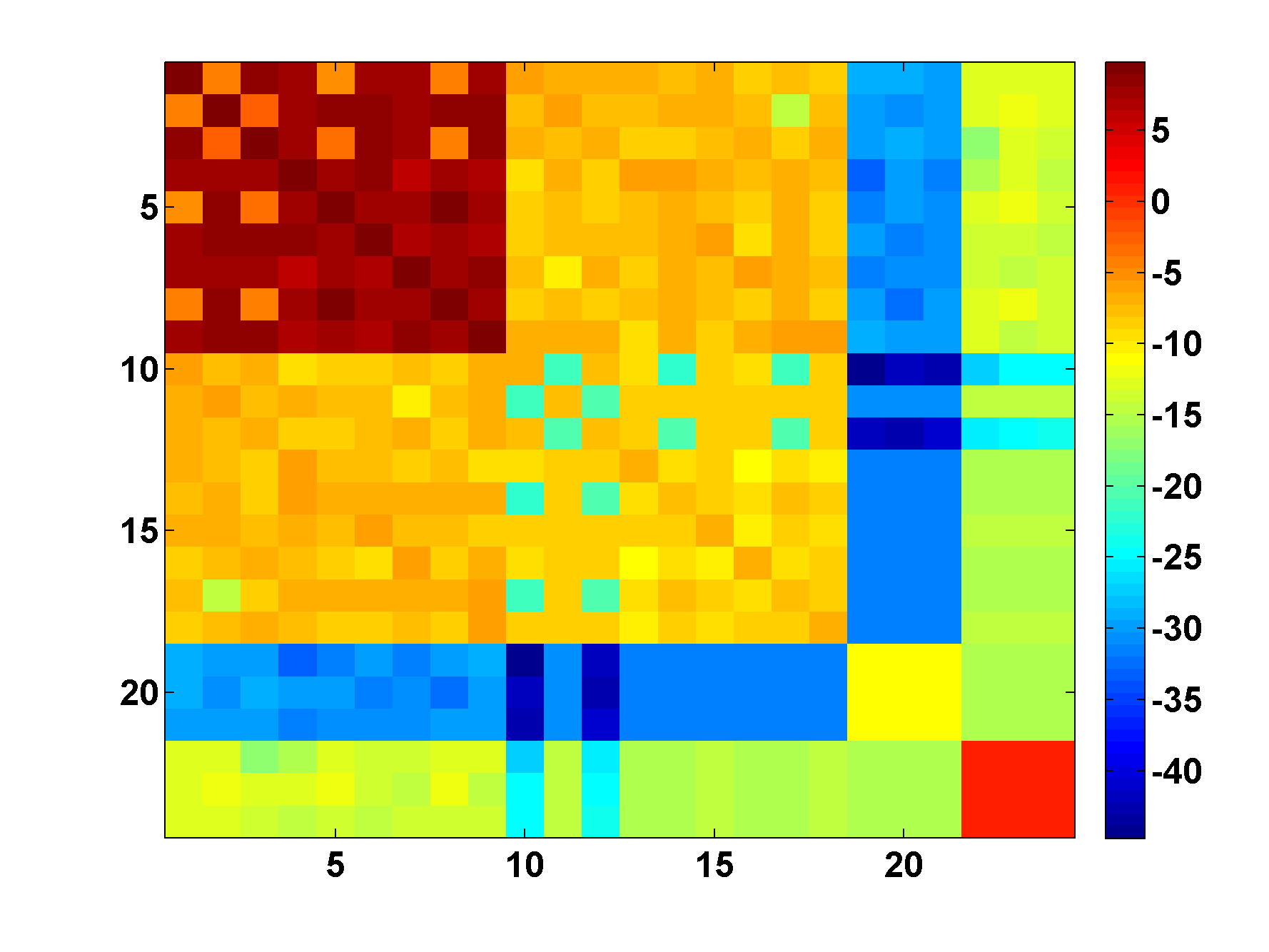}
\end{minipage}
}
\\
\subfloat[$P_{50}^-$.]{
\begin{minipage}[t]{0.25\textwidth}
\centering
\includegraphics[width=1.0\textwidth]{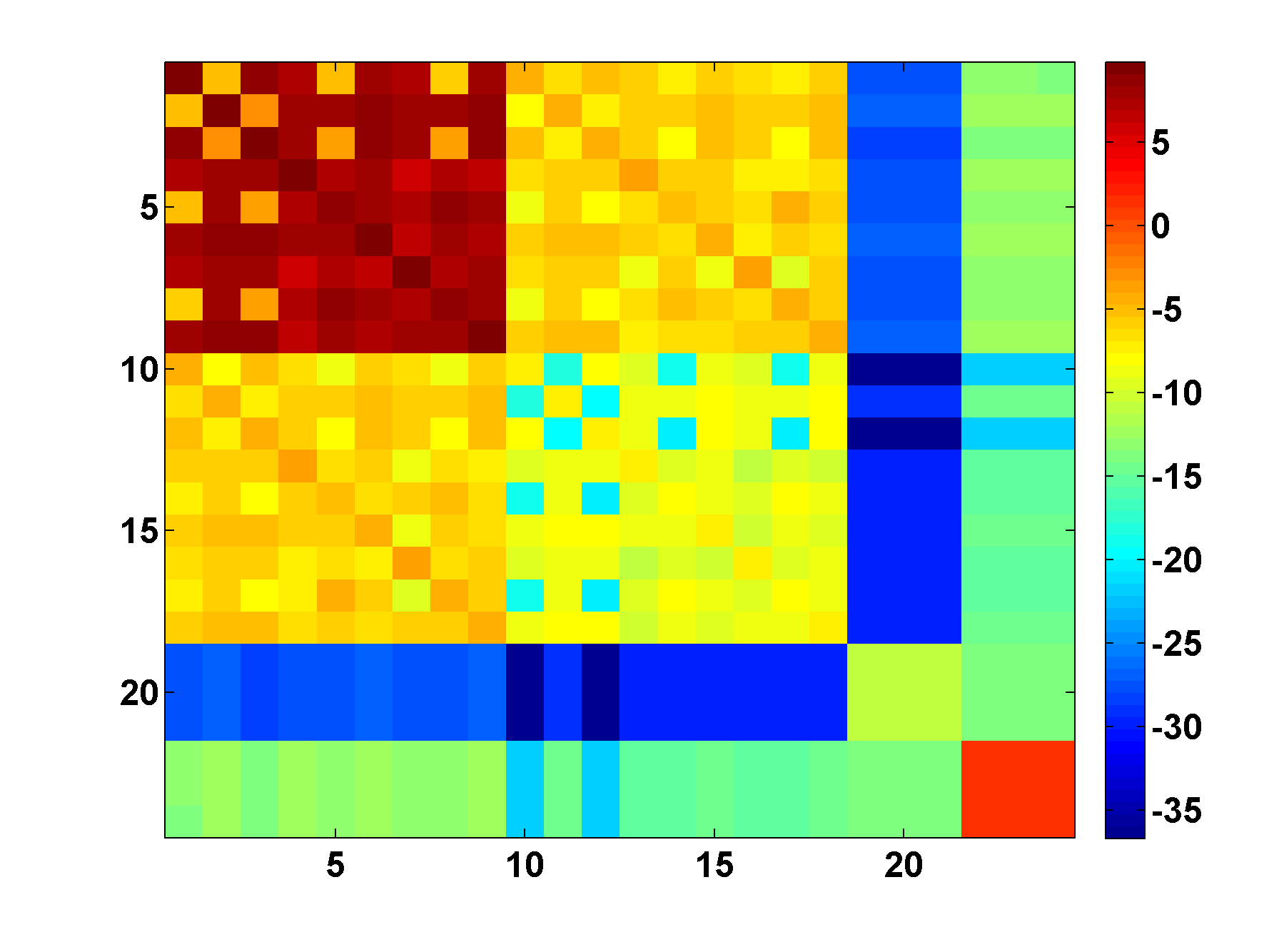}
\end{minipage}
}
\caption{ \label{fig:Pm}  A priori covariance matrices $P_k^-$ at different
steps. The absolute value of each component of the covariance matrix is represented by a color. The color map indicates the magnitude
of each component in logarithmic scale $\ln (|P_k^-|)$. }
\end{figure}

We then apply our previously designed hybrid extended Kalman filter to these measurements. The progress of the Kalman filter
can be characterized by looking at the uncertainty propagation. Fig.~\ref{fig:Pm} shows a priori covariance matrices at different
steps $k=\{1,2,5,10,50\}$. The absolute value of each component of the covariance matrix is represented by a color. The color map indicates the magnitude
of each component in logarithmic scale. The first covariance matrix $P_1^-$ is diagonal, since we do not assume prior knowledge of the
off-diagonal components. As the filter runs, the off-diagonal components emerge automatically from the system model, which can be seen
from Fig.~\ref{fig:Pm}. The initial uncertainties are relatively large. In fact, the initial positions are known only to about $20\,$km through
the deep space network (DSN). The uncertainties are significantly reduced after taking into account the precise inter-spacecraft measurements.
However, the uncertainties are not being reduced continuously. Instead, they stay roughly at the same level. This is because there are only
18 measurements at each step, whereas there are 24 variables in the state vector to be determined. There is not enough information to
precisely determine every variable in the state vector.

\begin{figure}[htbp]
\centering
\subfloat[$P_1^+$.]{
\begin{minipage}[t]{0.25\textwidth}
\centering
\includegraphics[width=1.0\textwidth]{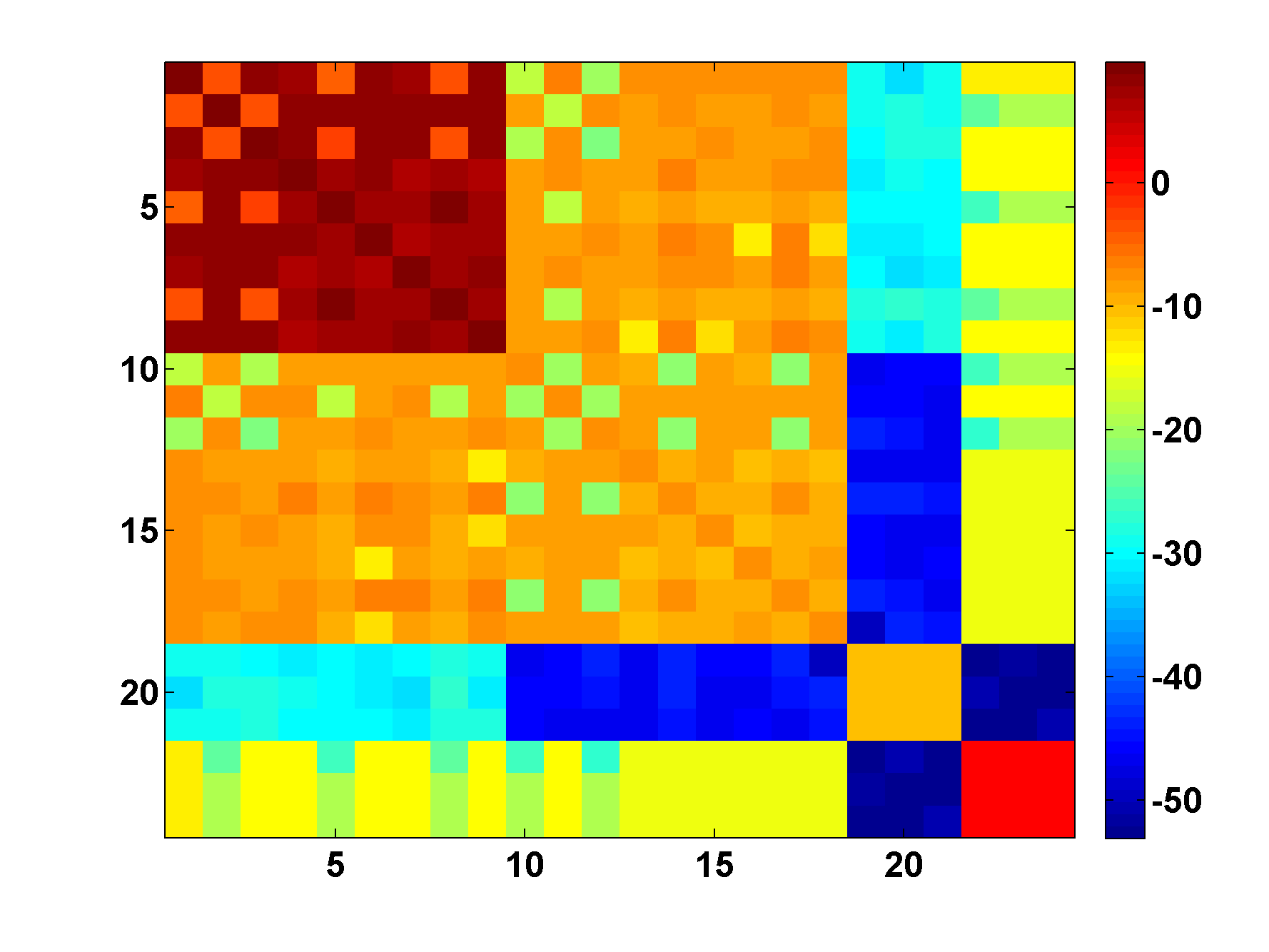}
\end{minipage}
}
\subfloat[$P_2^+$.]{
\begin{minipage}[t]{0.25\textwidth}
\centering
\includegraphics[width=1.0\textwidth]{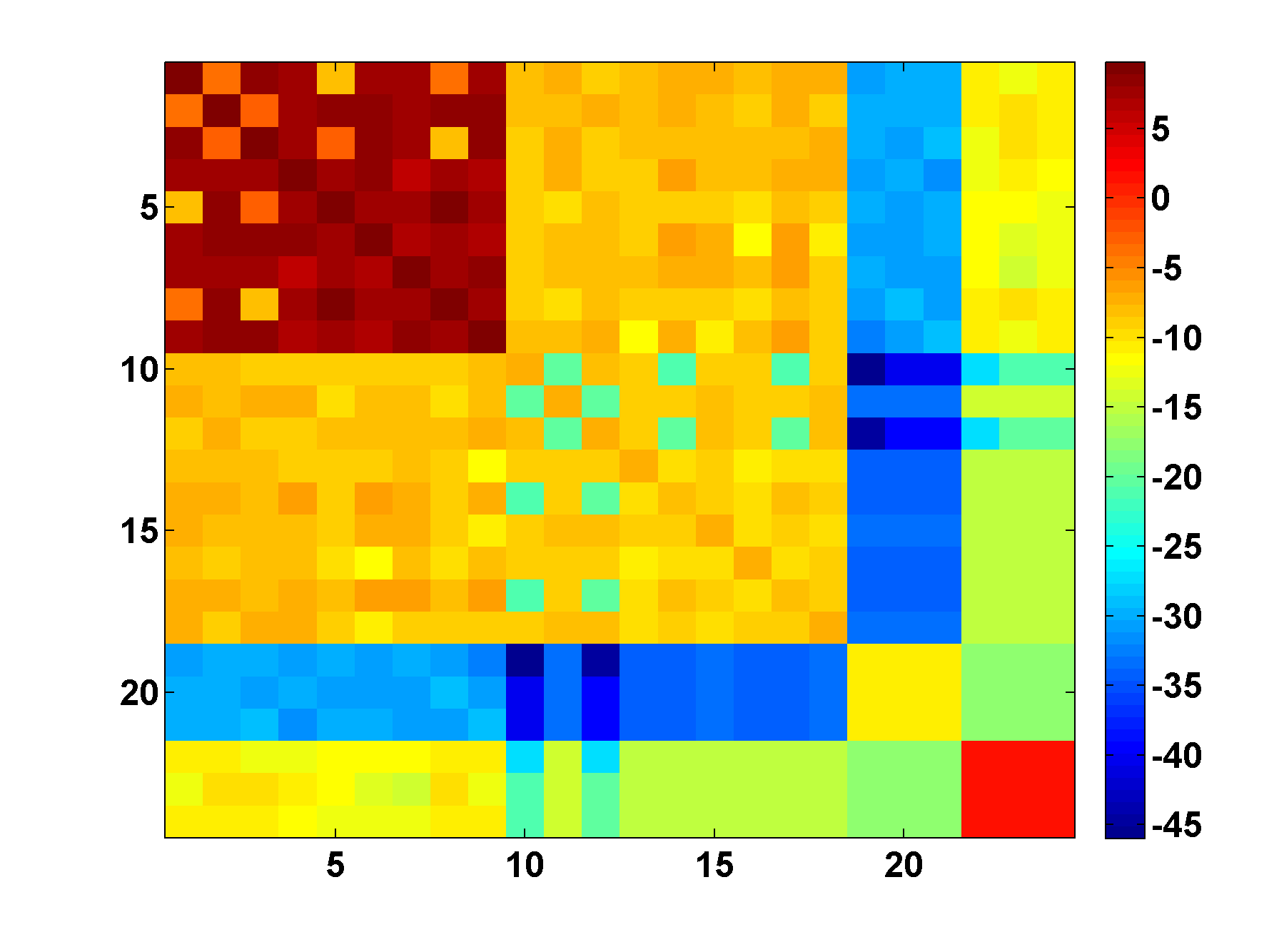}
\end{minipage}
}
\\
\subfloat[$P_5^+$.]{
\begin{minipage}[t]{0.25\textwidth}
\centering
\includegraphics[width=1.0\textwidth]{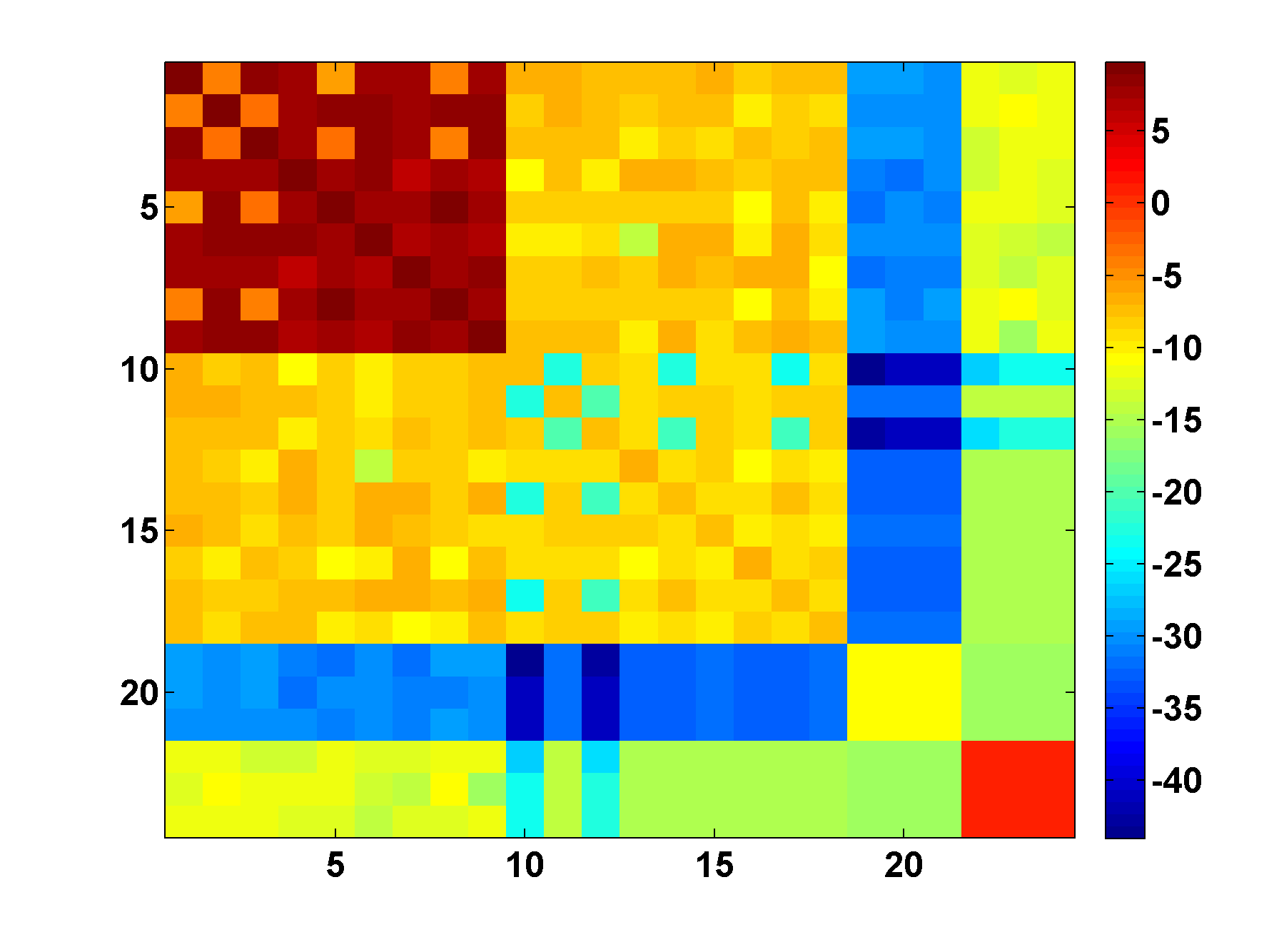}
\end{minipage}
}
\subfloat[$P_{10}^+$.]{
\begin{minipage}[t]{0.25\textwidth}
\centering
\includegraphics[width=1.0\textwidth]{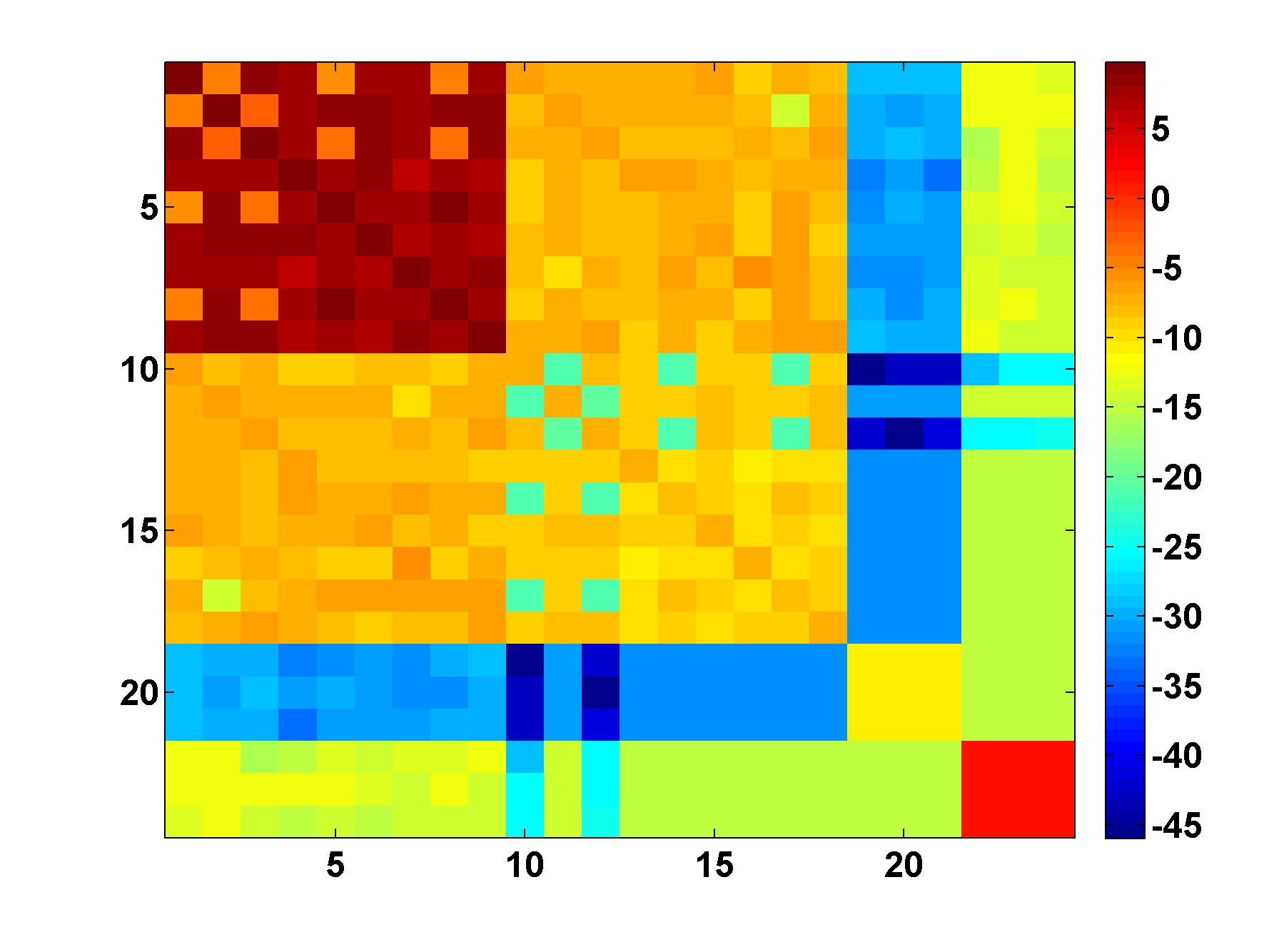}
\end{minipage}
}
\\
\subfloat[$P_{50}^+$.]{
\begin{minipage}[t]{0.25\textwidth}
\centering
\includegraphics[width=1.0\textwidth]{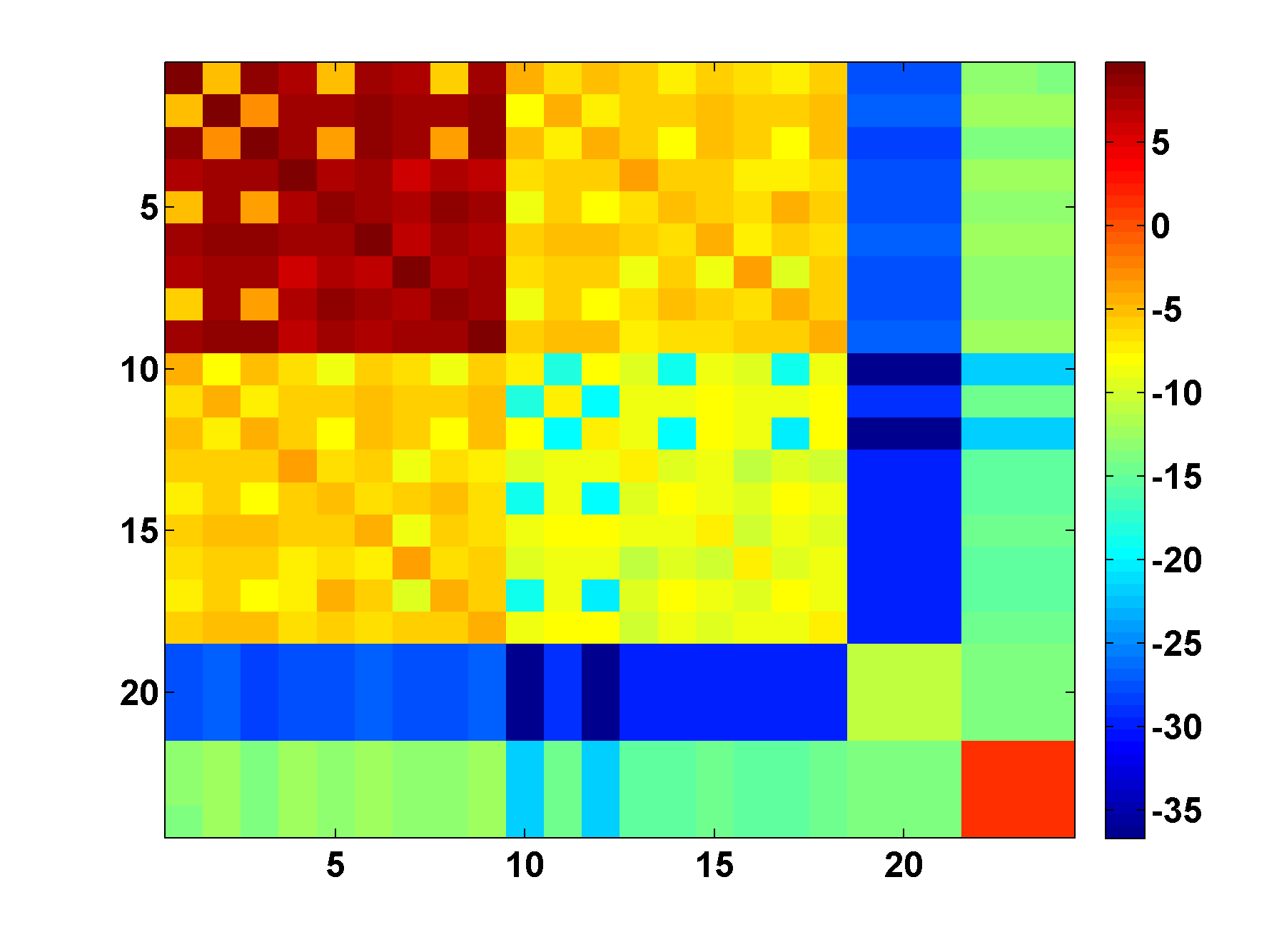}
\end{minipage}
}
\caption{ \label{fig:Pp} A posteriori matrices $P_k^+$ at different steps. The absolute value of each component
of the covariance matrix is represented by a color. The color map indicates the magnitude of each component
in logarithmic scale.}
\end{figure}

Similar behavior can be observed from the a posteriori covariance matrices in Fig.~\ref{fig:Pp}, where
the uncertainties also roughly stay at the same level. By comparing Fig.~\ref{fig:Pp} with Fig.~\ref{fig:Pm},
we find that the uncertainties are only slightly reduced from $P_k^-$ to $P_k^+$ with the help of the measurements
$y_k$. This is again because there are fewer measurements than variables in the state vector. Seemingly, this
hybrid extended Kalman filter does not work well. However, our aim is actually to reduce the noise in the measurements.
Let us denote the Kalman filter estimate of the measurements $y_k$ as $\hat{y}_k$, which can be calculated from the a
posteriori state vector as follows
\begin{eqnarray}
\hat{y}_k = H_k \hat{x}_k^+.
\end{eqnarray}

\begin{figure}[htbp]
\centering
\subfloat[Step 1.]{
\begin{minipage}[t]{0.25\textwidth}
\centering
\includegraphics[width=1.0\textwidth]{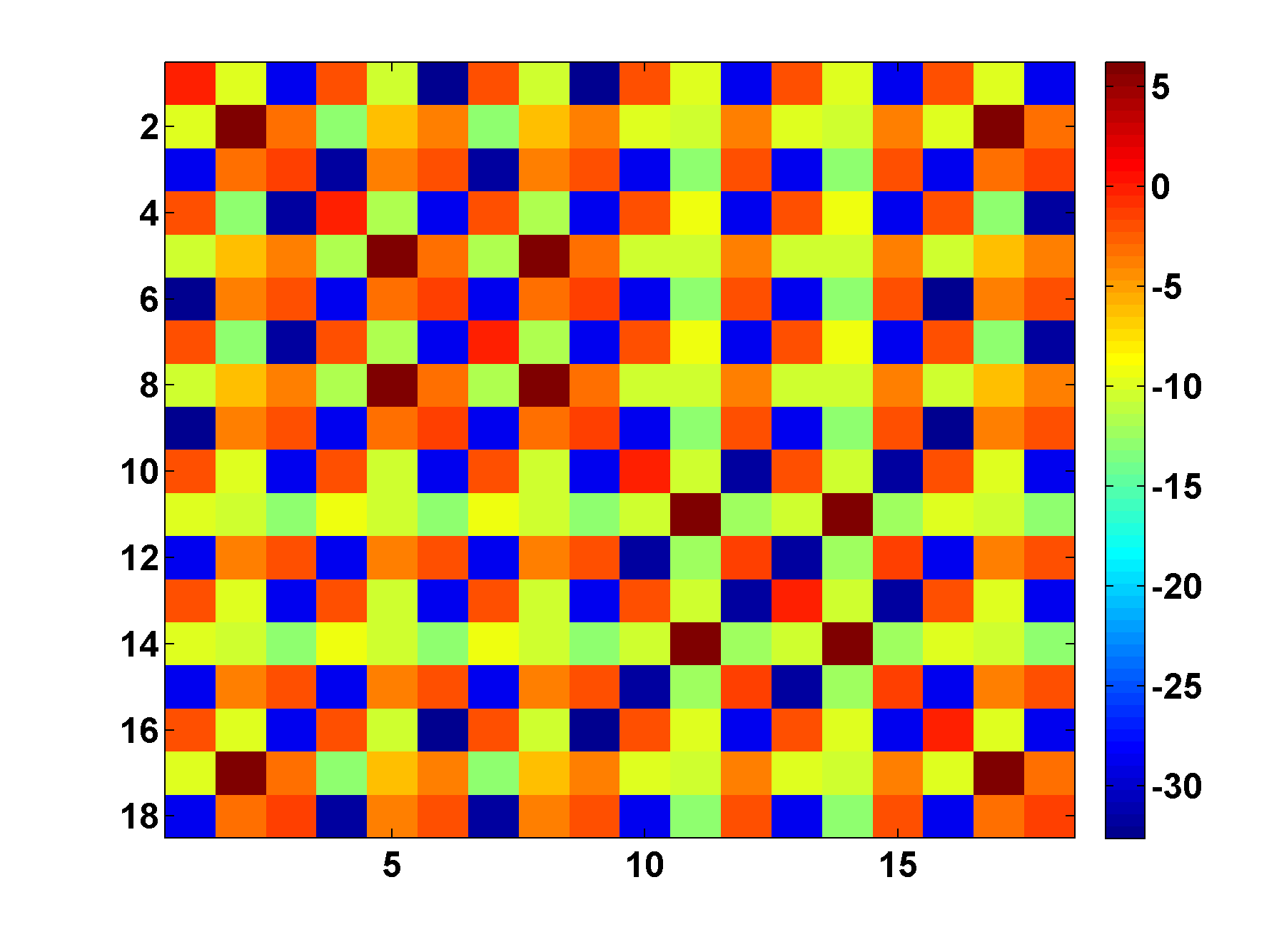}
\end{minipage}
}
\subfloat[Step 2.]{
\begin{minipage}[t]{0.25\textwidth}
\centering
\includegraphics[width=1.0\textwidth]{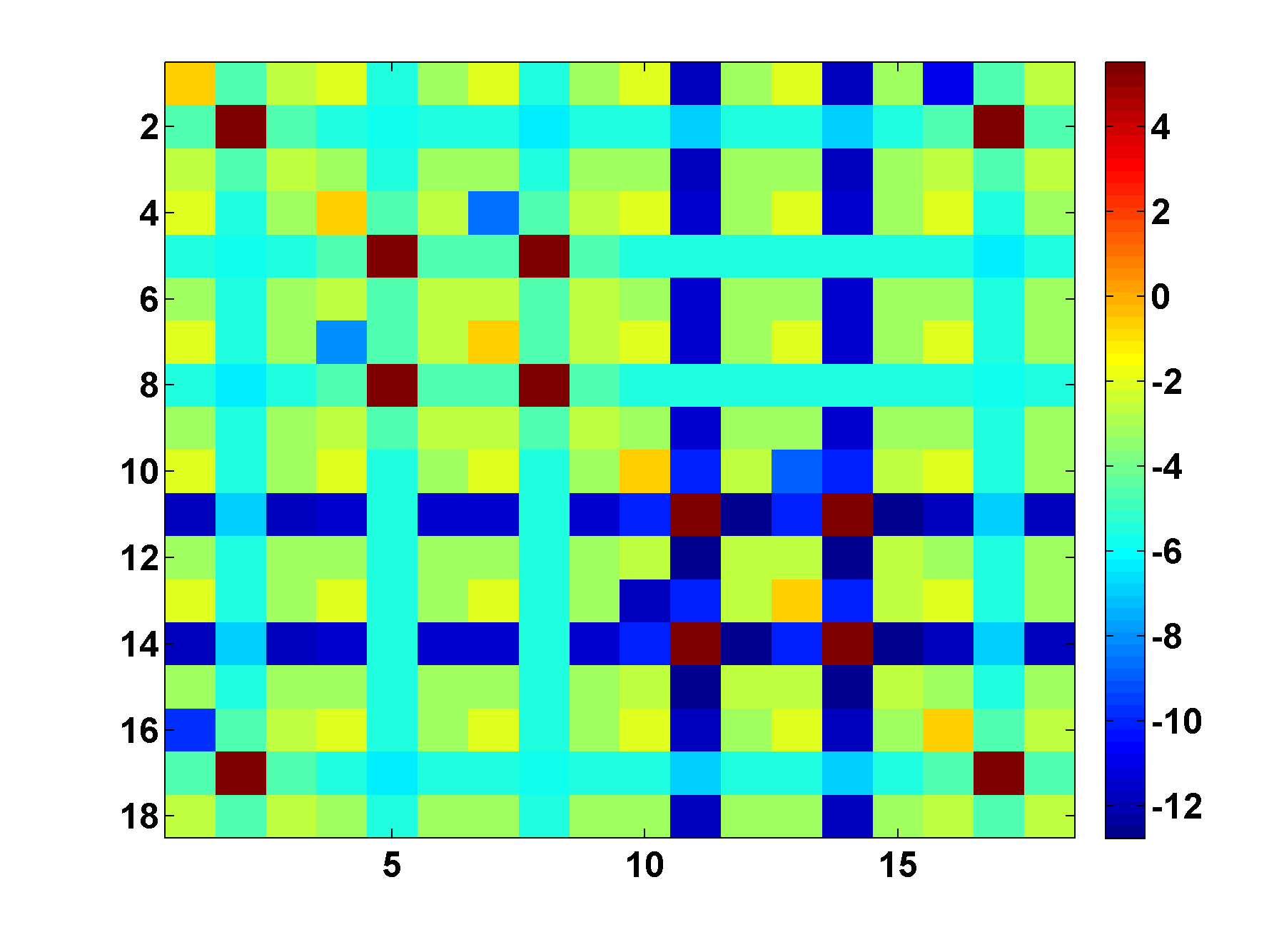}
\end{minipage}
}
\\
\subfloat[Step 5.]{
\begin{minipage}[t]{0.25\textwidth}
\centering
\includegraphics[width=1.0\textwidth]{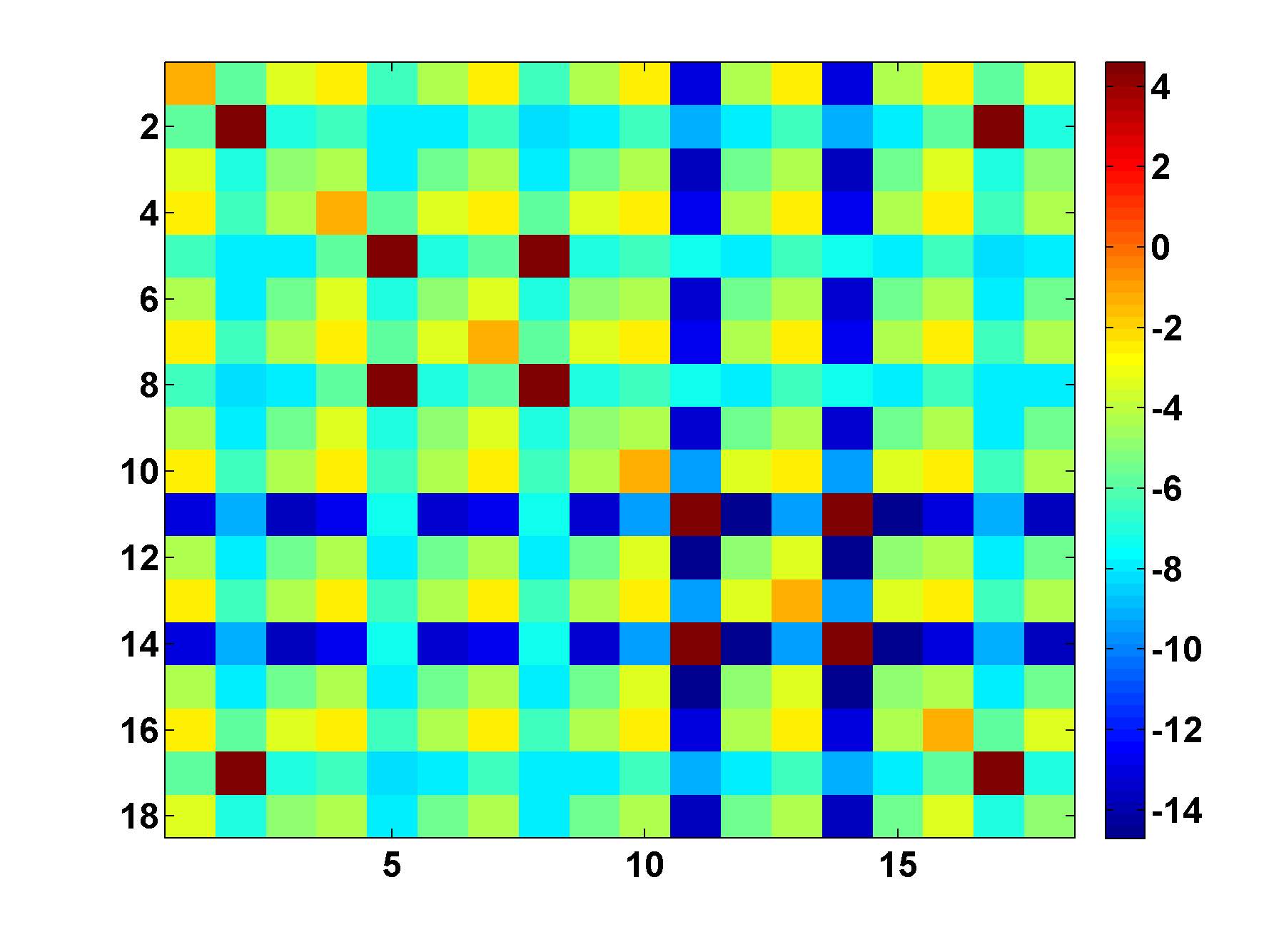}
\end{minipage}
}
\subfloat[Step 10.]{
\begin{minipage}[t]{0.25\textwidth}
\centering
\includegraphics[width=1.0\textwidth]{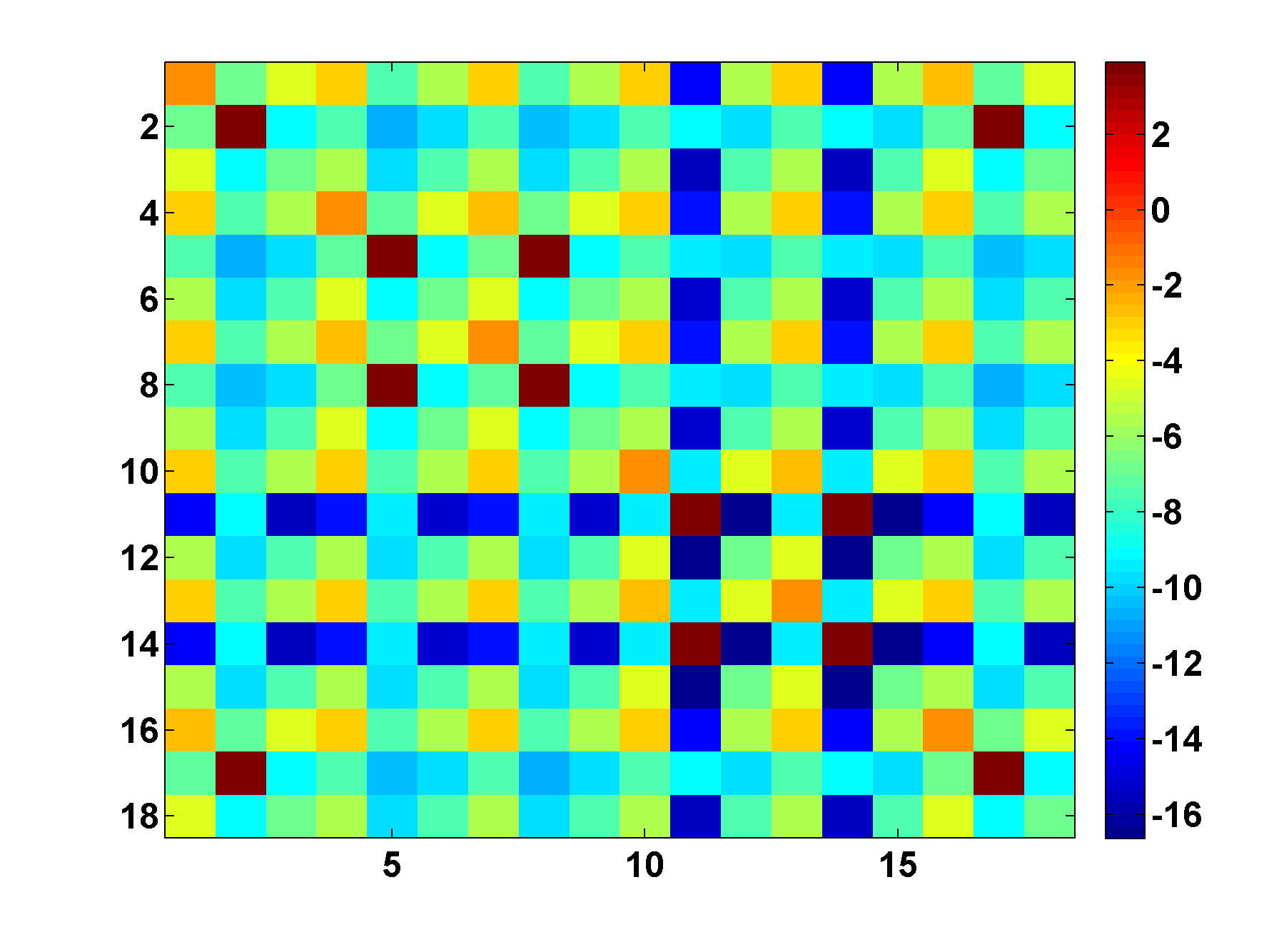}
\end{minipage}
}
\\
\subfloat[Step 50.]{
\begin{minipage}[t]{0.25\textwidth}
\centering
\includegraphics[width=1.0\textwidth]{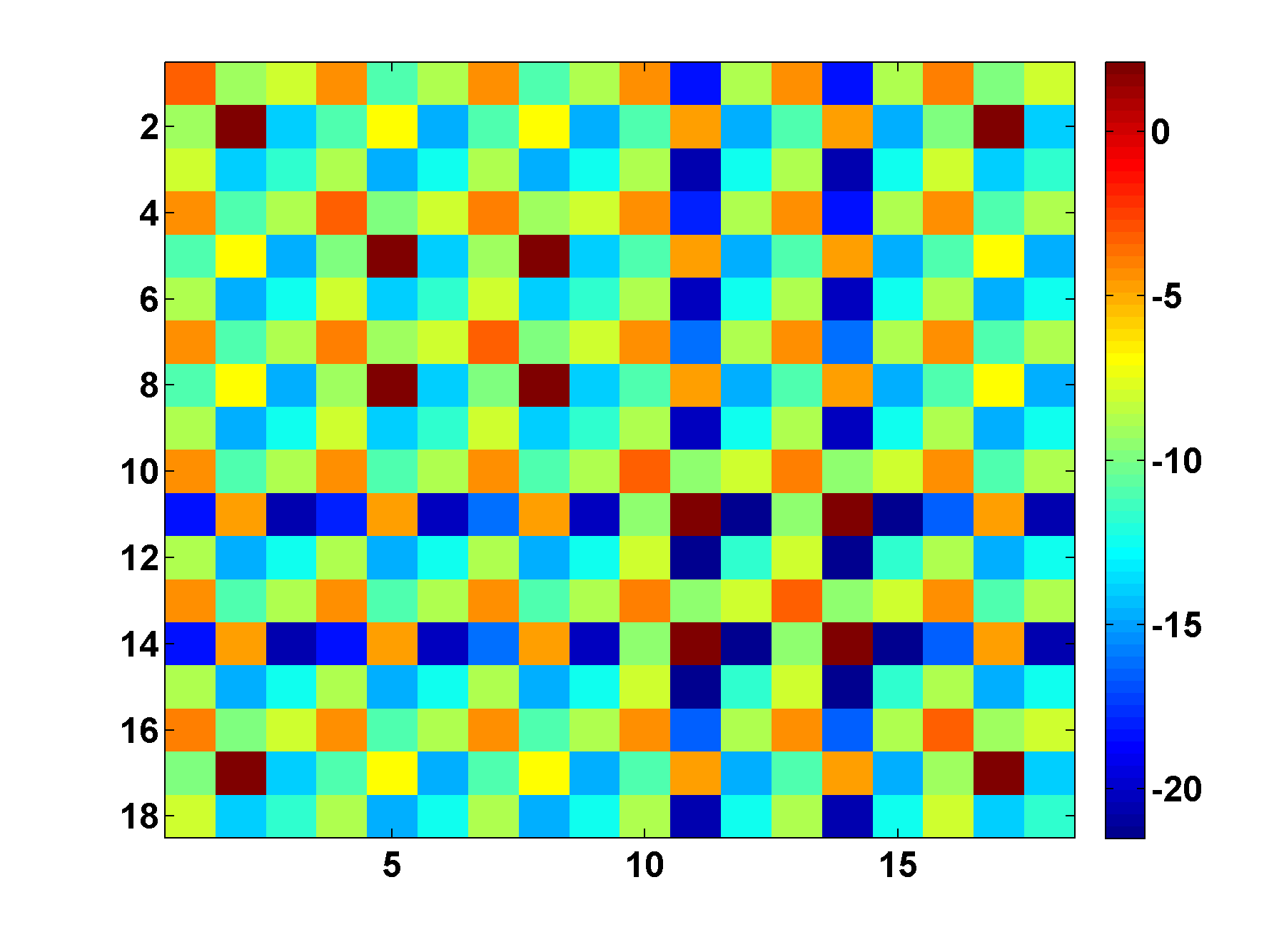}
\end{minipage}
}
\caption{ \label{fig:HppH} The estimation error of the measurements, $H_k P_k^+ H_k^T$ at different steps.
The absolute value of each component is represented by a color. The color map indicates the magnitude of each component
in logarithmic scale.}
\end{figure}

It is easy to show that the estimation error of $y_k$ can be expressed as $H_k P_k^+ H_k^T$,
which is shown in Fig.~\ref{fig:HppH}. Notice that the color bar shrinks with steps.
It is apparent that estimation errors of the measurements are significantly reduced by
the hybrid-extended Kalman filter. This is what is expected, since the number of the measurements $y_k$
is now the same as the number of variables $\hat{y}_k$ to be estimated in this case.

\begin{figure}[htbp]
\includegraphics[width=0.5\textwidth]{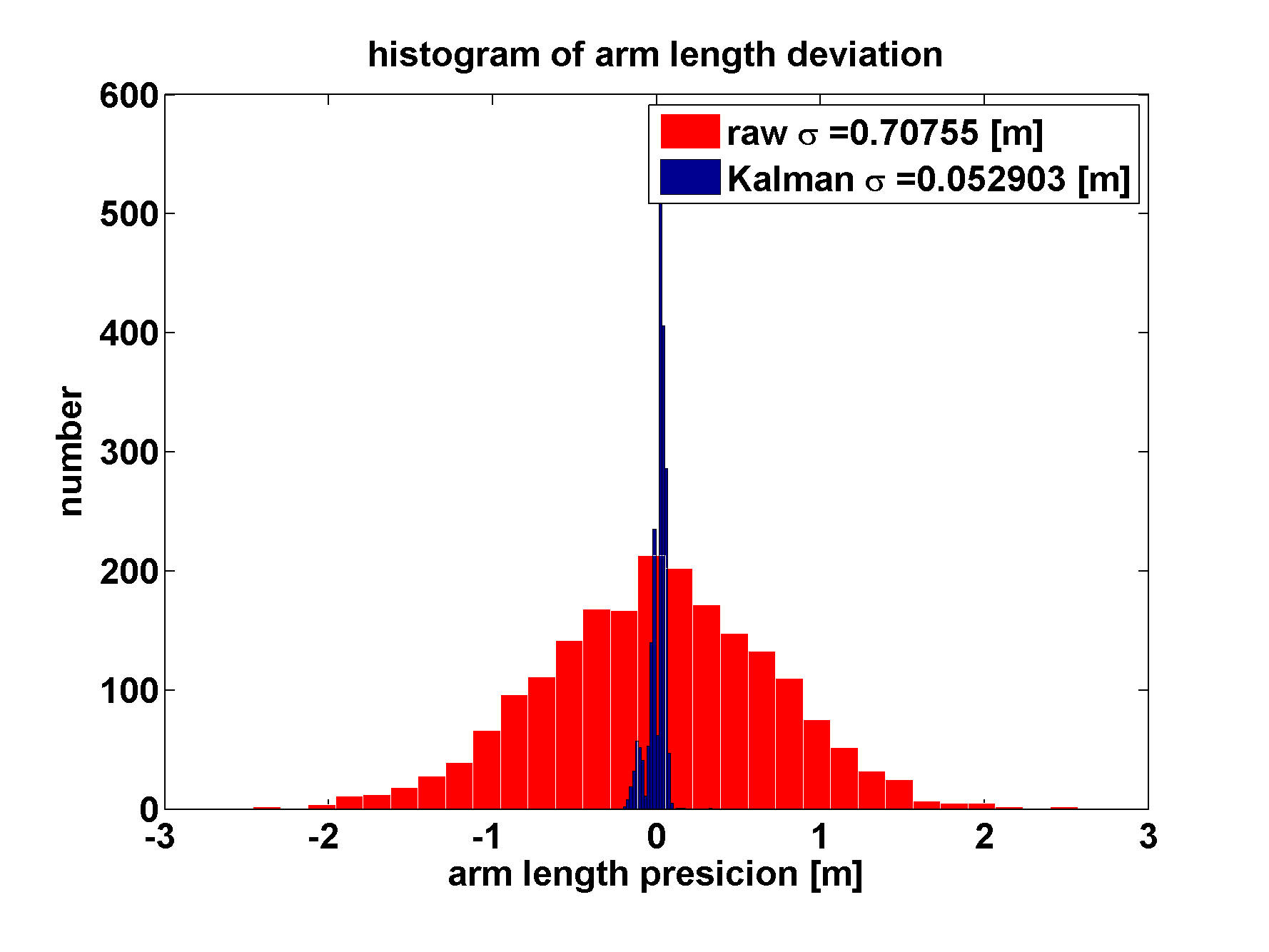}
\caption{ \label{fig:L} Histograms of errors of raw armlength measurements and Kalman filter estimates, where the deviations of both
raw arm-length measurements (excluding the initial clock bias) and the Kalman filter estimates from the true armlengths are
shown.}
\end{figure}

\begin{figure}[htbp]
\centering
\subfloat[]{
\begin{minipage}[t]{0.5\textwidth}
\centering
\includegraphics[width=1.0\textwidth]{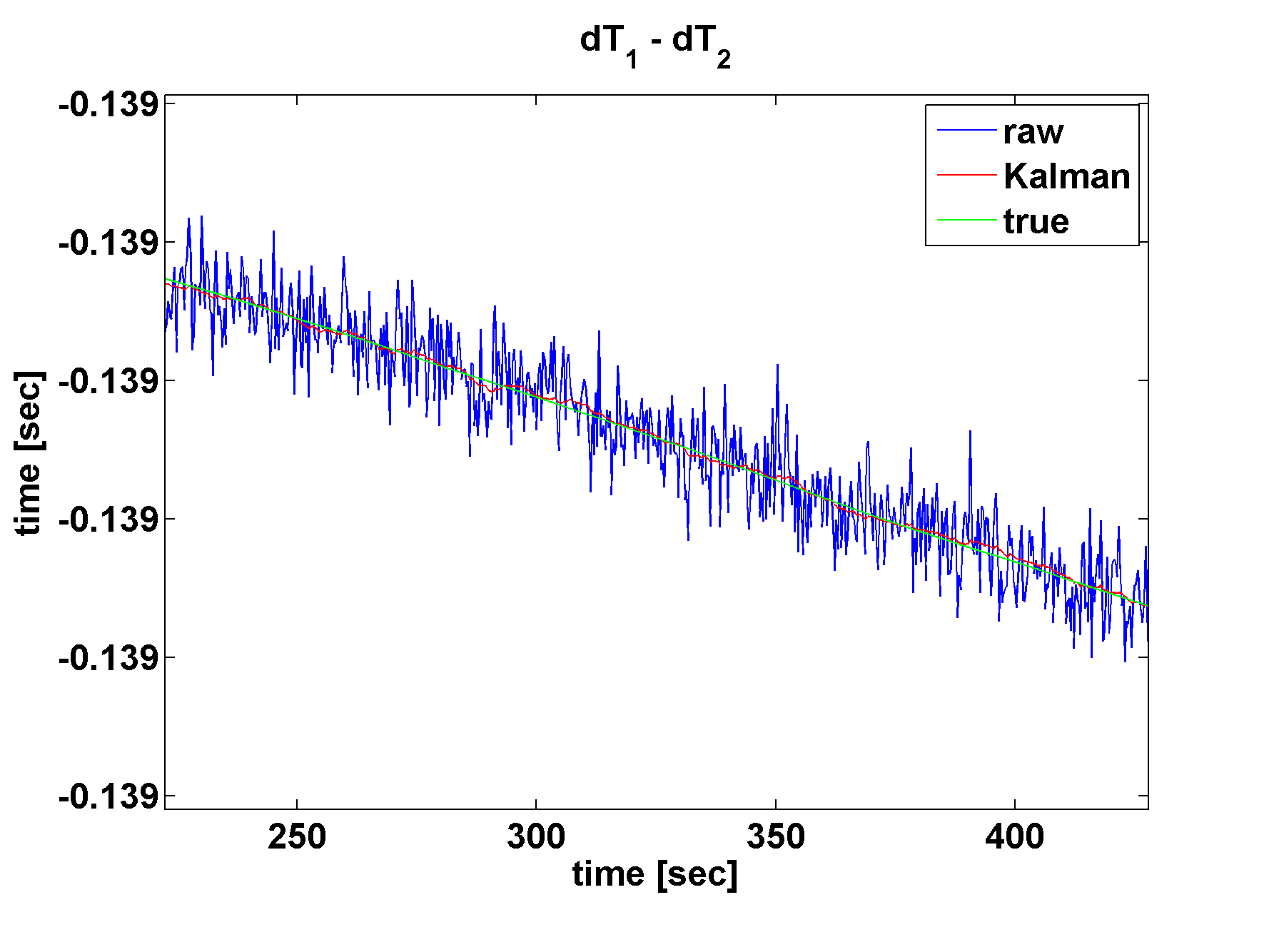}
\end{minipage}
}
\\
\subfloat[]{
\begin{minipage}[t]{0.5\textwidth}
\centering
\includegraphics[width=1.0\textwidth]{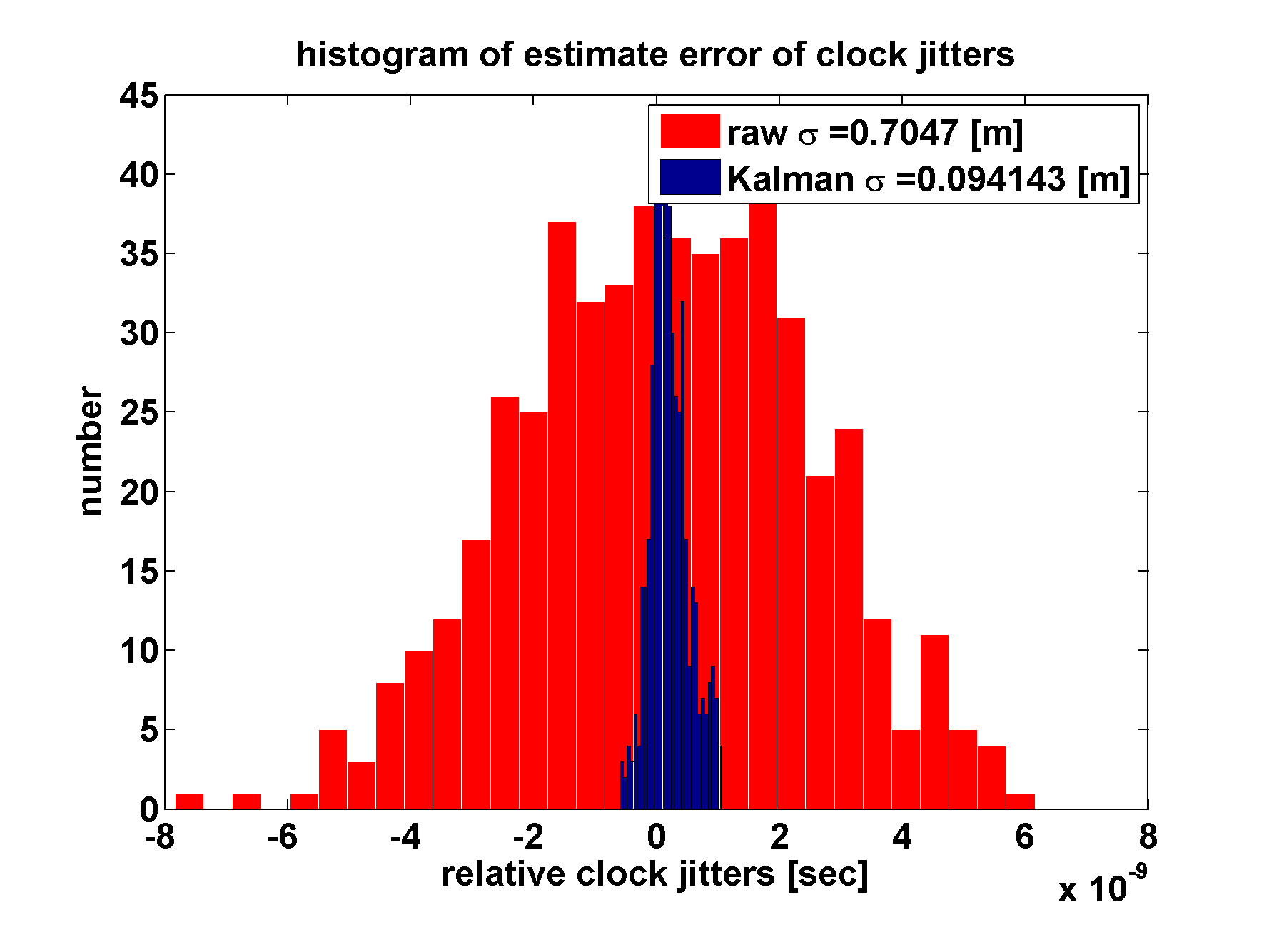}
\end{minipage}
}
\caption{ \label{fig:dT} Plots of relative clock jitter and biases. Fig.~(a) shows typical results of estimates of relative clock jitters and biases.
Fig.~(b) shows the deviations of the raw measurements and the Kalman filter estimates from the true values in histograms.
Notice that the standard deviations in the legend have
been converted to equivalent lengths.
}
\end{figure}

\begin{figure}[htbp]
\includegraphics[width=0.5\textwidth]{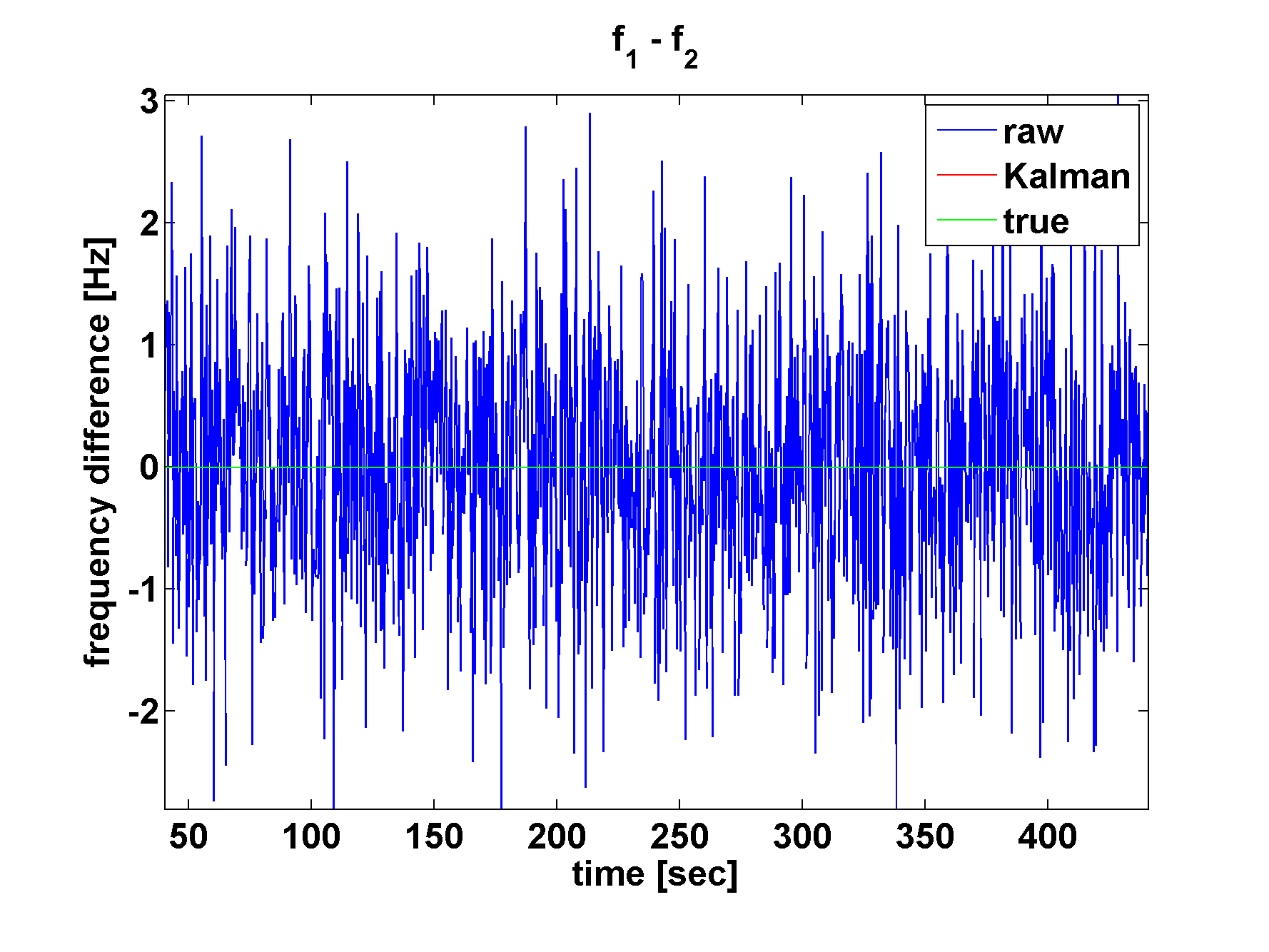}
\caption{ \label{fig:df1} The raw measurements, Kalman filter estimates
and the true values of frequency differences between the USO in S/C 1 and the USO in S/C 2. The Kalman filter estimates
are so good that they overlap with the true values in the figure.  }
\end{figure}

\begin{figure}[htbp]
\includegraphics[width=0.5\textwidth]{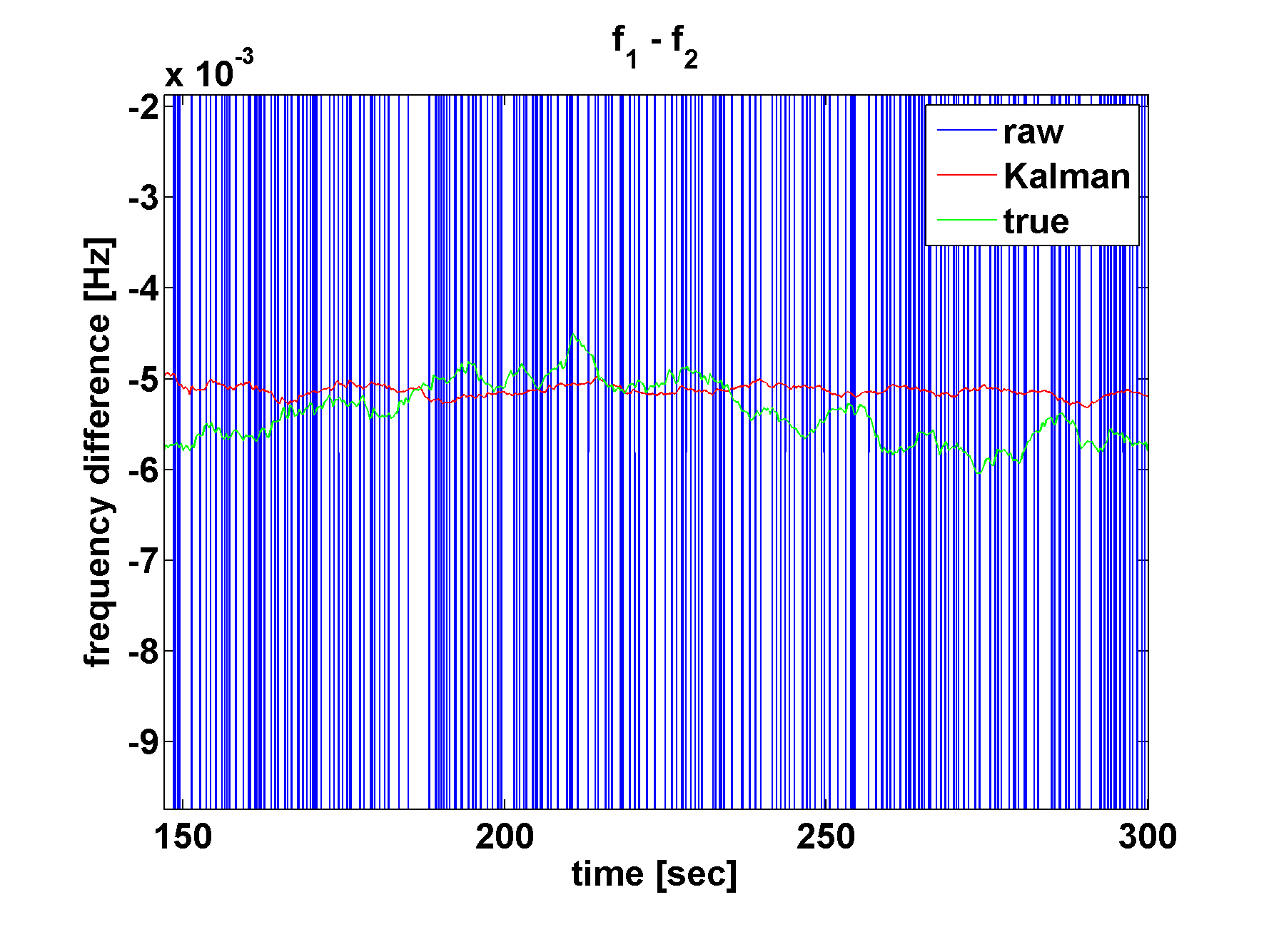}
\caption{ \label{fig:df2} A zoomed-in plot of Fig.~\ref{fig:df1}. The true USO frequency
differences and the Kalman filter estimates can clearly be seen in this figure.  }
\end{figure}

\begin{figure}[htbp]
\includegraphics[width=0.5\textwidth]{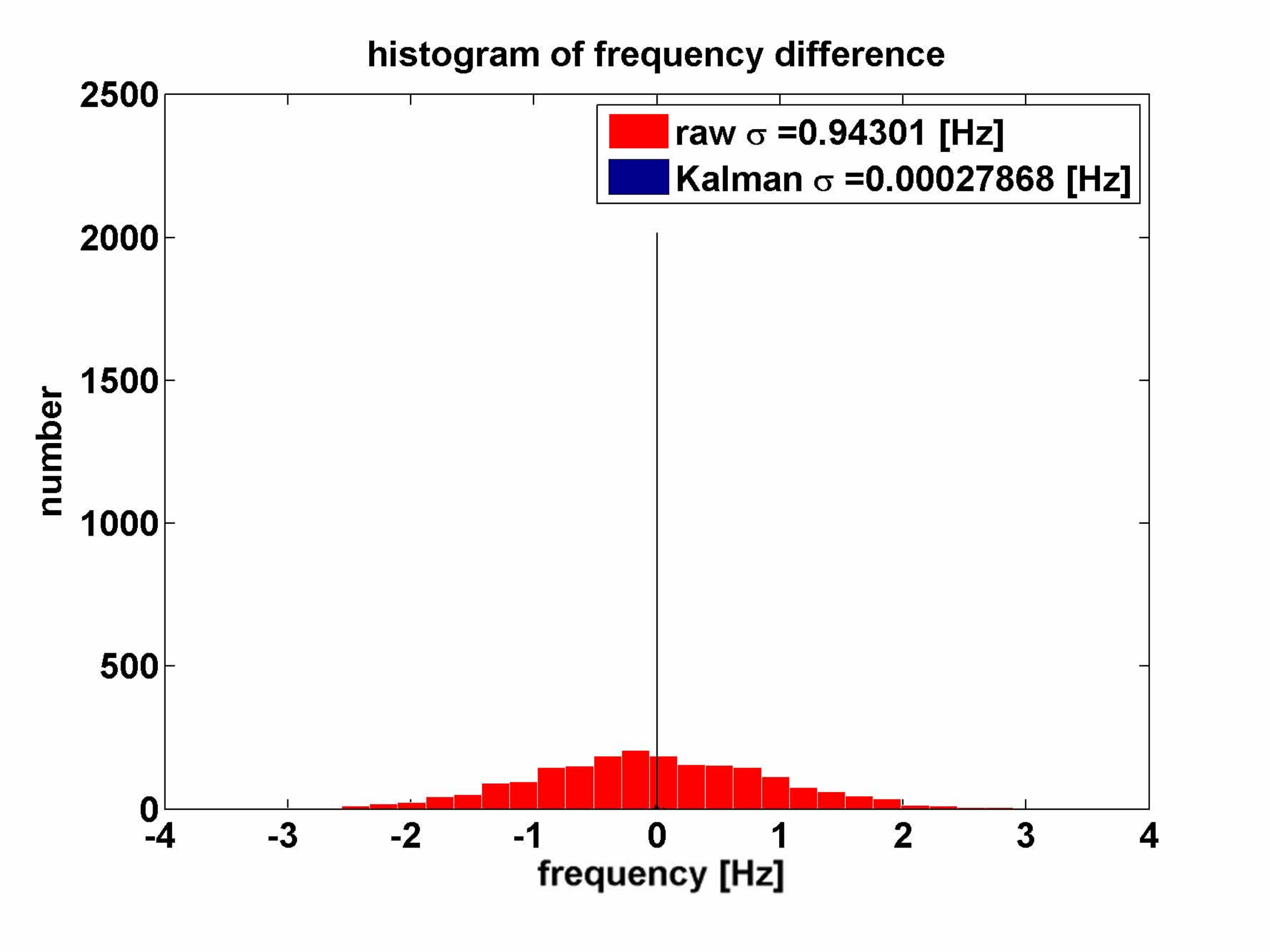}
\caption{ \label{fig:df3} The histograms of the deviations
of the raw measurements and the Kalman filter estimates from the true values. }
\end{figure}

Detailed simulation results are shown in Figs.~\ref{fig:L}, \ref{fig:dT}, \ref{fig:df1}, \ref{fig:df2} and \ref{fig:df3}.
Fig.~\ref{fig:L} exhibits histograms of errors of raw armlength measurements and Kalman filter estimates, where the deviations of both
raw arm-length measurements (excluding the initial clock bias) and the Kalman filter estimates from the true armlengths are
shown. The designed Kalman filter has not only decoupled the arm lengths from the clock biases better than 1 m rms, but also reduced the measurement
noise by more than one order of magnitude to the centimeter level. This precise arm-length knowledge is necessary to
allow excellent performance of TDI techniques, which subsequently permits optimal extractions of the science information
from the measurement data.

Fig.~\ref{fig:dT} (a) shows typical results of estimates of relative clock jitters and biases,
where the blue curve stands for the raw measurements, the green curve exhibits the true time difference between the clock in S/C 1 and S/C 2, the red
curve plots the Kalman filter estimates of the clock time differences. It is clear from the figure that the Kalman filter estimates resemble the true
values quit well. Fig.~\ref{fig:dT} (b) shows the deviations of the raw measurements and the Kalman filter estimates from the true values in histograms.
Notice that the standard deviations in the legend have been converted to equivalent lengths. It is apparent
that the designed Kalman filter has reduced the measurement noise by about an order of magnitude. These accurate clock jitter estimates
enable us to correct the clock jitters in the postprocessing step. Hence, it potentially allows us to use slightly poorer
clocks, yet still achieving the same sensitivity. This would potentially help reduce the cost of the mission.

Fig.~\ref{fig:df1} shows the raw measurements, Kalman filter estimates
and the true values of frequency differences between the USO in S/C 1 and the USO in S/C 2. The Kalman filter estimates
are so good that they overlap with the true values. Fig.~\ref{fig:df2} is a zoomed-in plot of Fig.~\ref{fig:df1}. The true USO frequency
differences and the Kalman filter estimates can clearly be seen in this figure. Fig.~\ref{fig:df3} shows the histograms of the deviations
of the raw measurements and the Kalman filter estimates from the true values. With the help of the designed Kalman filter,
the measurement noise has been reduced by 3-4 orders of magnitude. Frequency jitters are directly related
to the first differential of the clock drifts. Therefore, such precise estimates of the USO frequency differences
will allow a very accurate tracking of the relative clock drifts.


\section{Summary}
\label{sec:Sum}

We have modeled LISA inter-spacecraft measurements and designed a hybrid-extended Kalman filter to process
the raw measurement data.  In the designed Kalman filter model, there are 24 variables in the state vector
and 18 variables in the measurement vector. Therefore, (i) the state vector, in principle, cannot be fully determined
from the measurements, which is one of the major differences from the global positioning system (GPS)~\cite{GPS,Mohinder08}
tracking problem, where the number of measurements is larger than the number of unknowns in the state vector.
Other important differences from GPS are: (ii) the position and the time of the emitter S/C are unknown (more
specifically, they are also to be estimated). Therefore, the measurements associated with each laser link are
functions of the receiver S/C at the current time and the emitter S/C at a previous time. This gives rise to
the ``causality" problem that the inference of current S/C depends also on future measurements.
(iii) The measurements recorded on different S/C are unsynchronized, and contaminated by different unknown clock jitter.
These differences make the first stage of LISA data processing much more challenging.

This paper presents a major step towards a high fidelity end-to-end simulation of the entire
LISA data processing chain. We have identified the problems, established the framework and formulated
inter-spacecraft measurements that are crucial to the first stage data processing.
Two important effects have not yet been included in the current simulation: (i) the time delay is only
partly simulated. In the simulation, the ranging measurements consist of the time delay and other noise.
However, the dependence of inter-spacecraft measurements on the emitter S/C at a delayed time is simulated
as that at the current time. Therefore, the effects of the `causality' problem and the Sagnac
differential delay do not present in this simulation. These issues are being investigated in
our follow-on work \cite{Wang14b}, where we find out that it is more appropriate to simulate
these effects in the full-relativistic framework. (ii) The clock jitter is not included in the
recording time yet, but only in the measurements. This effect is included in our follow-on
work \cite{Wang14b}.

The current simulation shows that the hybrid-extended Kalman filter can well
decouple the arm lengths from the clock biases and significantly improve the relative measurements, such as
arm lengths, relative clock jitters and relative frequency jitters etc. However, the absolute variables in the
state vector cannot be determined accurately. These variables include the absolute positions and velocities of the
spacecraft, the absolute clock drifts and the absolute frequency drifts. This is mainly due to the fact that
only the differences are measured and the number of measurements is lower than the number of variables in the state vector.

It can be better understood by taking a closer look at the measurement equations~\ref{eq:R}, \ref{eq:D}
and \ref{eq:C}. In fact, only the relative positions $\sqrt{(x_j-x_i)^2+(y_j-y_i)^2+(z_j-z_i)^2}$
and the relative longitudinal velocities $(\vec{v}_j-\vec{v}_i)\cdot \hat{n}_{ij}$ appear in the measurements.
Neither absolute positions nor absolute velocities are directly measured. Thus, it is impossible to fully constrain
the entire LISA configuration only with these inter-spacecraft measurements. The clock jitters only appear in Eq.~\ref{eq:R}
in the form of $\delta T_j - \delta T_i$, which means the common clock drifts are undetermined. The relative USO frequency
jitters $\delta f_j - \delta f_i$ are measured in Eq.~\ref{eq:C}. The absolute USO frequency jitters $\delta f_j$ appear
in Eq.~\ref{eq:D}. However, $\delta f_j / f_j^\textrm{nom}$ is far less than 1, hence Eq.~\ref{eq:D} can provide only very limited
information about $\delta f_j$. As a result, the absolute USO frequency jitters $\delta f_j$ are poorly determined.

\appendix


\section{A proof of the optimality}

In the Kalman filter derivation, the Kalman gain $K_k$ is chosen such that the estimation error $\textrm{tr}(P_k^+)$ in the state vector
is minimized. However, in the LISA case we are interested in reducing the noise in the measured variables rather than reducing the uncertainties in the
state vector; Hence, the optimal filter in this case should minimize the estimation error in the measurements $y_k$.

In this appendix, we prove that minimizing the estimation error in the state vector $x_k$ is equivalent to minimizing the estimation error
in $y_k$ to the linear order. As shown in previous sections, the estimation error in $y_k$ is $\textrm{tr}(H_k P_k^+ H_k^T)$ in the linearized
model. To minimize the trace of this covariance matrix, we have
\begin{widetext}
\begin{eqnarray}
\frac{\partial [\textrm{tr}(H_k P_k^+ H_k^T)]}{\partial K_k} &=& \frac{\partial [\textrm{tr}(H_k^T H_k P_k^+ )]}{\partial K_k} \nonumber \\
&=& \frac{\partial \{\textrm{tr}[H_k^T H_k (I - K_k H_k) P^-_k (I - K_k H_k)^T + H_k^T H_k K_k V_k K_k ]\}}{\partial K_k} \nonumber \\
&=& 0
\end{eqnarray}
To be concise, we omit the step index $k$ and use the subscripts for the component indices.
\begin{eqnarray}
&& \frac{\partial \{\textrm{tr}[H^T H (I - K H) P^- (I - K H)^T]\}}{\partial K} \nonumber \\
&=& \frac{\partial \{\textrm{tr}[H^T_{ni} H_{ij} (I_{jl} - K_{jk} H_{kl}) P^-_{lm} (I - K H)^T_{mn}]\}}{\partial K_{ab}} \nonumber \\
&=& \frac{\partial \{\textrm{tr}[H_{in} H_{ij} (I_{jl} - K_{jk} H_{kl}) P^-_{lm} (I_{nm} - K_{nc} H_{cm})]\}}{\partial K_{ab}} \nonumber \\
&=& H_{in} H_{ij} ( - \delta_{aj}\delta_{bk} H_{kl}) P^-_{lm} (I_{nm} - K_{nc} H_{cm}) + H_{in} H_{ij} (I_{jl} - K_{jk} H_{kl}) P^-_{lm} ( - \delta_{an}\delta_{bc} H_{cm}) \nonumber \\
&=& - H^T_{ai} H_{in}(I-KH)_{nm}P^{-T}_{ml}H^T_{lb} - H^T_{ai} H_{ij}(I-KH)_{jl}P^-_{lm}H^T_{mb} \nonumber \\
&=& - 2 H^T H(I-KH)P^-H^T,
\end{eqnarray}
\end{widetext}
where we have adopted Einstein summation convention and used the fact that $P^+$ is symmetric. Similarly, we have
\begin{eqnarray}
\frac{\partial \{\textrm{tr}( H^T H K V K )\}}{\partial K} = 2 H^T H K V.
\end{eqnarray}
By putting back the step index $k$, we have
\begin{eqnarray}
0 &=& \frac{\partial [\textrm{tr}(H_k P_k^+ H_k^T)]}{\partial K_k} \nonumber \\
&=& 2 H^T_k H_k [ K_k V_k - (I-K_k h_k)P^-_k H^T_k ].
\end{eqnarray}
The Kalman gain is then solved as follows
\begin{eqnarray}
K_k = P^-_k H^T_k (H_k P_k^- H_k^T + V_k)^{-1},
\end{eqnarray}
which is the same as what we have used.


\begin{acknowledgements}

Y.W. would like to thank A. R{\"u}diger and M. Hewitson for useful suggestions,
and S. Barke for providing a figure.
The authors are partially supported by DFG Grant No. SFB/TR 7 Gravitational Wave
Astronomy and the DLR (Deutsches Zentrum fu\"r Luft- und
Raumfahrt). The authors would like to thank the German Research Foundation for funding
the Cluster of Excellence QUEST-Center for Quantum Engineering and
Space-Time Research.
\end{acknowledgements}

\end{document}